\documentclass[12pt,a4paper,english]{article}

\usepackage[margin=1.1in]{geometry}
\newcommand{\ignore}[1]{}
\usepackage{changepage}

\usepackage{adjustbox}
\usepackage[graphicx]{realboxes}
\usepackage{longtable}
\usepackage[utf8]{inputenc}
\usepackage{titleps}
\usepackage{authblk}
\usepackage{setspace}
\usepackage{sectsty}
\usepackage{titling}
\usepackage{titletoc}
\usepackage{pifont}
\usepackage{ragged2e}

\usepackage{csquotes}

\usepackage{lscape}
\usepackage[bf,sf,justification=centering,font=small]{caption}
\usepackage{subcaption}
\usepackage{float}
\usepackage{rotating}
\usepackage{multicol}
\usepackage{threeparttable}
\usepackage{multirow}
\usepackage{comment}
\usepackage{booktabs}
\usepackage{lipsum}
\usepackage{tikz}
\usepackage{graphicx}

\usetikzlibrary{decorations.pathreplacing,positioning, arrows.meta}
\definecolor{myLightGray}{RGB}{191,191,191}
\definecolor{myGray}{RGB}{160,160,160}
\definecolor{myDarkGray}{RGB}{144,144,144}
\definecolor{myDarkRed}{RGB}{167,114,115}
\definecolor{myRed}{RGB}{255,58,70}
\definecolor{myGreen}{RGB}{0,255,71}
\usepackage{caption}
\usepackage{subcaption}

\usepackage{natbib}

\usepackage{multibib}

\usepackage{amsfonts}
\usepackage{amsmath}
\usepackage{amssymb}
\usepackage{amsthm}
\usepackage{commath}
\usepackage{bbm}
\usepackage[nottoc]{tocbibind}
\usepackage{mathtools}

\usepackage{nicefrac}

\sectionfont{\large}
\subsectionfont{\normalsize\mdseries\bf}

\newpagestyle{online}{%
  \sethead{}{}{\bf For Online Publication}%
  \setfoot{}{\thepage}{}%
  \setheadrule{0pt}%
  \setfootrule{0pt}%
}

\usepackage[pdfborder={0 0 0 [0 0]}, colorlinks=FALSE, urlcolor=magenta,	 citecolor=red, filecolor=magenta, bookmarks=TRUE, pdftitle= {DMRP_XXXXXX},	 pdfauthor={DMM}]{hyperref}

\hypersetup{
    colorlinks = {true},
    citecolor={blue},
    urlcolor={blue},
    linkcolor = {blue}
    }

\begin{document}

\title{Free Public Transport: \\ More Jobs without Environmental Damage?~}
\makeatletter\let\Title\@title\makeatother
%Does Free Public Transport Impact Economic and Environmental Outcomes? Evidence from Brazil

\title{\Large \textbf{\Title}%
\thanks{Rodrigues: Sao Paulo School of Economics - FGV, e-mail: \href{mailto:mateus.rodrigues@fgv.edu.br}{mateus.rodrigues@fgv.edu.br}. Da Mata: Sao Paulo School of Economics - FGV, e-mail: \href{mailto:daniel.damata@fgv.br}{daniel.damata@fgv.br}. Possebom: Sao Paulo School of Economics - FGV, e-mail: \href{mailto:vitor.possebom@fgv.edu.br}{vitor.possebom@fgv.br}. We are grateful to Rafael Araujo, Bladimir Carrillo, Lucas Finamor and Sophie Mathes for invaluable comments and suggestions. The usual disclaimer applies. Declarations of interest: none.
}}

\author{{ \large Mateus Rodrigues}
\and { \large Daniel Da Mata}
\and { \large Vitor Possebom}
}
\bigskip
\date{\normalsize \today}
\maketitle

\begin{center} 
\large \textbf{Latest version available \href{https://sites.google.com/site/danielddamata/research}{here}.}
\end{center} 

\begin{abstract}
\noindent

%100 words

We study the effects of a free-fare transport policy implemented by Brazilian localities on employment and greenhouse gas emissions. Using a staggered difference-in-differences approach, we find that fare-free transit increases employment by 3.2\% and reduces emissions by 4.1\%, indicating that transport policies can decouple economic activity from environmental damage. Our results are driven by workers transitioning from higher-emission to lower-emission sectors instead of being driven by a decline in private transportation use. Cost-benefit analyses suggest that the costly policy only presents net benefits after considering the tax inflows of the increased economic activity and the benefits of reduced carbon emissions.

\bigskip
\bigskip

\bigskip


\end{abstract}

\onehalfspacing

\setlength\abovedisplayskip{5pt}
\setlength\belowdisplayskip{5pt}
\thispagestyle{empty}\clearpage
\pagenumbering{arabic}%

%%%%%%%%%%%%%%%%%%%%%%%%%%%%%%%%%%%%%%%%
\section{Introduction}\label{sec:introduction}
%%%%%%%%%%%%%%%%%%%%%%%%%%%%%%%%%%%%%%%%

The implications of transport policies for the economy and the environment have been a recurring theme in policy debates. While transport policies may boost mobility with positive consequences for economic activity, the mobility surge and increased economic activity may also be accompanied by more greenhouse gas (GHG) emissions---the primary cause of climate change and a major contributor to environmental damage. As a result, policymakers are interested in policies that, by altering behavioral responses, are able to decouple economic activity from GHG emissions.\footnote{In this paper, we define ``decoupling'' as the act of increasing economic activity while decreasing environmental damage.} While the electrification of buses and the use of biofuels have been prominent strategies to mitigate emissions, evidence of how alternative policies---such as free transport passes---can help decoupling economic outcomes from emissions is rather scarce \citep{jaramillo2022transport}.

What are the economic and environmental effects of free public transport? Can free public transport decouple economic activity from environmental damage? We study these questions in the context of Brazil. More specifically, we investigate the effects of a transport policy that provides (permanent) \textit{free} bus fares to all users in all routes to boost mobility in Brazilian municipalities.\footnote{Municipalities in Brazil are local autonomous political-administrative entities roughly equivalent to U.S. counties and have the power to implement urban transport policies.} Using a staggered difference-in-difference approach and a number of data sources, our empirical analysis explores the program's rollout across Brazilian municipalities. We focus on two outcomes: employment and GHG emissions. We then connect these two outcomes to assess whether the policy can decouple employment growth from emissions.

An empirical analysis of free transport policies is important because the effects of such policies on employment and emissions are conceptually ambiguous. Free-fare bus rides are likely to facilitate job search and positively influence the employment of job-seekers directly affected by the policy (\citealp{Franklin2017}). However, as the increased employment may come from the displacement of existing workers, employment levels will rise if the policy also affects labor demand---e.g., by influencing the expansion of incumbent firms or creating new ones \citep{Tsivanidis2024}---or the matching between workers and firms \citep{Agrawal2024}. Therefore, free transport policies do not mechanically increase employment. Similarly, these policies do not automatically decrease emissions. While free public transport may potentially decrease emissions by encouraging commuters to switch from cars to buses, variations in public fares might also increase emissions by spurring broader economic changes and increasing economic activity. Besides, free transport policies may generate over-consumption of bus rides, which increase emissions. 

Our results indicate that the free public transport policy increases employment by 3.2\% and reduces GHG emissions by 4.1\%. Therefore, our findings show that this policy decouples employment from environmental damage. Notice that decoupling can be \textit{absolute}, when economic activity increases and emissions do not, or \textit{relative}, when economic activity increases more than the rise in emissions. Therefore, we find evidence of absolute decoupling. Importantly, an event-study analysis suggests that the effect on employment is persistent over time while the effect on GHG emissions may be short-lived.

We also analyze possible mechanisms driving these results. When we analyze the potential mechanisms through which the policy affects labor markets, we find that the policy is associated with the creation of new firms, which is consistent with the increased employment we observe. As for the channels related to decreased GHG emissions, we find that the results are not driven by a decline in private transport use. Instead, our results are driven by changes in job composition, with individuals transitioning from higher-emission to lower-emission sectors. Specifically, we observe workers changing from agricultural to urban jobs. 

We also carry out a cost-benefit analysis to further understand the implications of this (costly) policy funded by the government budget. Our back-of-the-envelope calculations consider the government expenditures due to larger subsidies and two groups of benefits: a ``national” benefit in the form of fiscal externalities---as the increase in observed employment generates a tax inflow to the economy---and a ``global” benefit in the form of reduced carbon emissions. Considering the national and global benefits, our calculations suggest that the policy presents net benefits. Importantly, the policy is only cost-effective when considering all benefits, such as the tax revenue accrued by the national government---whose transference to localities in Brazil is not compulsory---and the benefits of reduced carbon emissions---and there is no market where Brazilian localities are compensated for internalizing the external cost of emissions. Two implications from the analysis are that (i) the policy is not cost-effective when considering only the national benefits, and (ii) the take-up of such transport policies may be lower when there is no market compensating localities for the benefits of reducing a global externality.

Our setting is suitable for understanding the effects of free public transport for at least four reasons. First, policymakers are interested in how transport policies impact behavioral responses and environmental outcomes because transport is a key sector in the low-carbon transition. Second, there is an interest in policies that can promote decoupling and benefit the lower-income population. Third, free public transport policies can be implemented in many places because bus networks are central in developed and developing countries.\footnote{Bus networks account for a large share of ridership worldwide, and many cities worldwide have recently implemented free-fare public transport, such as Albuquerque, United States, and Tallinn, Estonia.} Fourth, Brazil offers rich data on emissions, employment, expenditures, and taxes, allowing us to analyze this policy's costs, benefits, and efficacy. Since the policy is costly, cost-benefit analyses are relevant in informing other contexts about this policy efficacy.

This paper contributes to four literatures. First, we connect to the literature on how transportation affects economic outcomes---see \cite{Duranton2020} for a survey. We also relate to studies investigating the role of transport in reducing spatial constraints in job search and spatial mismatch (e.g., \citealp{Phillips2014, Franklin2017, Banerjee2023}). Our results are consistent with studies highlighting that reduced commuting costs facilitate access to formal jobs (e.g., \citealp{Khanna2022, zarate2022spatial}). In a literature review, \cite{Gonzalez-Navarro2023} point out that intercity transportation improvements have strong effects on shifting labor out of agriculture in rural villages. We add by showing that (intracity) transport policies affect formal employment and sectoral change.\footnote{The shift in sector composition that we observe aligns with the broader literature on how shocks affect structural transformation (e.g., oil discoveries in \citealp{Cavalcanti2019} and property rights \citealp{adamopoulos2022land}). Since we find that a reduction in spatial frictions generates structural transformation, we also connect to the  macro development literature on the structural transformation out of agriculture---see \cite{Gollin2023} for a literature review.}

Second, we also connect to studies on the effects of free transport interventions. Existing studies focus on how temporary subsidies interfere with travel behavior (\citealp{Bull2021} and \citealp{Brough2022}.) and employment (\citealp{Brough2024}). Since the policy we analyze provides a permanent subsidy to all city residents, we add by analyzing aggregate effects on employment and environmental externalities.

Third, we connect to the literature on the environmental externalities of transport. In particular, we relate to the strand on the effects of policies on GHG emissions, which explores the effects of various transport policies, including subsidies \citep{qin2015designing}, congestion \citep{bharadwaj2017impact}, vehicle types \citep{wang2018automated, lin2021electric}, and transport mode choices \citep{gillingham2019tale, donna2021measuring, lin2021impact}. We contribute by showing that---given the high-dosage nature of zero-price policies---the environmental impacts can come from changing the composition of economic activity (i.e., structural transformation). 

Finally, this study relates to the literature on the determinants of GHG emissions, especially the strand on how public policies affect emissions. Within this strand, few papers investigate how policies affect decoupling.\footnote{See, for instance, the systemic review in \citet{Haberl2020}.} Understanding the relationship between policies and decoupling is relevant as anthropogenic GHG emissions reached unprecedented levels over the past decade \citep{ipcc2022climate}, with significant economic repercussions, including income losses \citep{burke2015global, kahn2021long}, spatial inequality \citep{cruz2024economic}, and shifts in economic activity \citep{desmet2015spatial, conte2021local}. Relative to that literature, this study is unique in studying whether free transport policies can decouple economic activity from emissions. The Intergovernmental Panel on Climate Change (IPCC) points out that free transport policy is a potentially relevant strategy for a low-carbon economy \citep{jaramillo2022transport}. We provide new empirical evidence indicating that this policy can increase employment while reducing emissions.

The remainder of this paper is organized as follows. Section \ref{sec:background} gives an overview of free public transport in Brazil. Section \ref{sec:data} presents the data. Section \ref{sec:empirical_strategy} discusses our empirical strategy. Section \ref{sec:results} presents the main results. Section \ref{sec:further_analyses} analyzes potential mechanisms and reports the cost-benefit analysis. Finally, Section \ref{sec:conclusion} concludes.

%%%%%%%%%%%%%%%%%%%%%%%%%%%%%%%%%%%%%%%%
\section{Background}\label{sec:background}
%%%%%%%%%%%%%%%%%%%%%%%%%%%%%%%%%%%%%%%%

%%%%%%%%%%%%%%%%%%%%%%%%%%%%%%%%%%%%%%%%
\subsection{Free Public Transport in Brazil}
%%%%%%%%%%%%%%%%%%%%%%%%%%%%%%%%%%%%%%%%

Brazil is a three-tiered federation, and municipal governments, the third tier of Brazil's federation, are charged with implementing urban transport policies.\footnote{Brazil has 26 states, a federal district, and 5,571 municipalities. Municipalities in Brazil are local autonomous political-administrative entities roughly equivalent to U.S. counties.} Municipalities can provide public transport within their jurisdiction via direct provision or concession agreements with private companies. In Brazil, public-private partnerships play an important role in urban transport, as most of the localities in the country provide public transport through concession agreements.

Brazil's public transport system mainly consists of buses and rail services, with buses being the dominant mode of transport---buses account for 85.7\% of all public transport journeys \citep{NTU2020}. Urban public transport in Brazil rarely funds itself with user fees; most typically, its funding comes from a combination of user fees and resources from the municipalities' budget. 

In the past decades, fare-free public transport of buses has emerged as an increasingly prevalent policy in Brazil, with many municipalities adopting it over time. In this paper, we study the effect of the \textit{universal free public transport} policy, when all users have unrestricted access to all bus routes every day without charge. There are partial fare-free public transport policies in the country, but we focus on the high-dosage universal fare-free policy because it is the one being increasingly adopted by local governments.\footnote{Examples of partial fare-free public transport policies in Brazilian municipalities include when (i) specific user categories are totally or partially exempt from bus fares (e.g., workers of shopping malls in São Luís and students in São Paulo), when (ii) users incur no charges on certain days of the week (e.g., free transport on Sundays in São Paulo), or when (iii) bus routes serve specific areas for free (e.g., buses serving informal settlements in Belo Horizonte).} Policymakers rationalize adopting fare-free public transport on equity and efficiency grounds. The usual argument is that the policy will boost the local economy, primarily benefiting the mobility of lower-income bus users even though the policy is not targeted specifically at the lower-income population. Universal free public transport funding comes from the municipality budget.\footnote{To fund free passes, localities use alternative funding sources beyond taxation, including advertising on buses, rental income from shops at terminals, and (a share of) traffic fines.}

\begin{figure}[htbp]
\begin{center}
\caption{Location of Municipalities Adopting Universal Free Public Transport}
\label{fig:map_policy_adoption}
\includegraphics[width=0.85\linewidth]{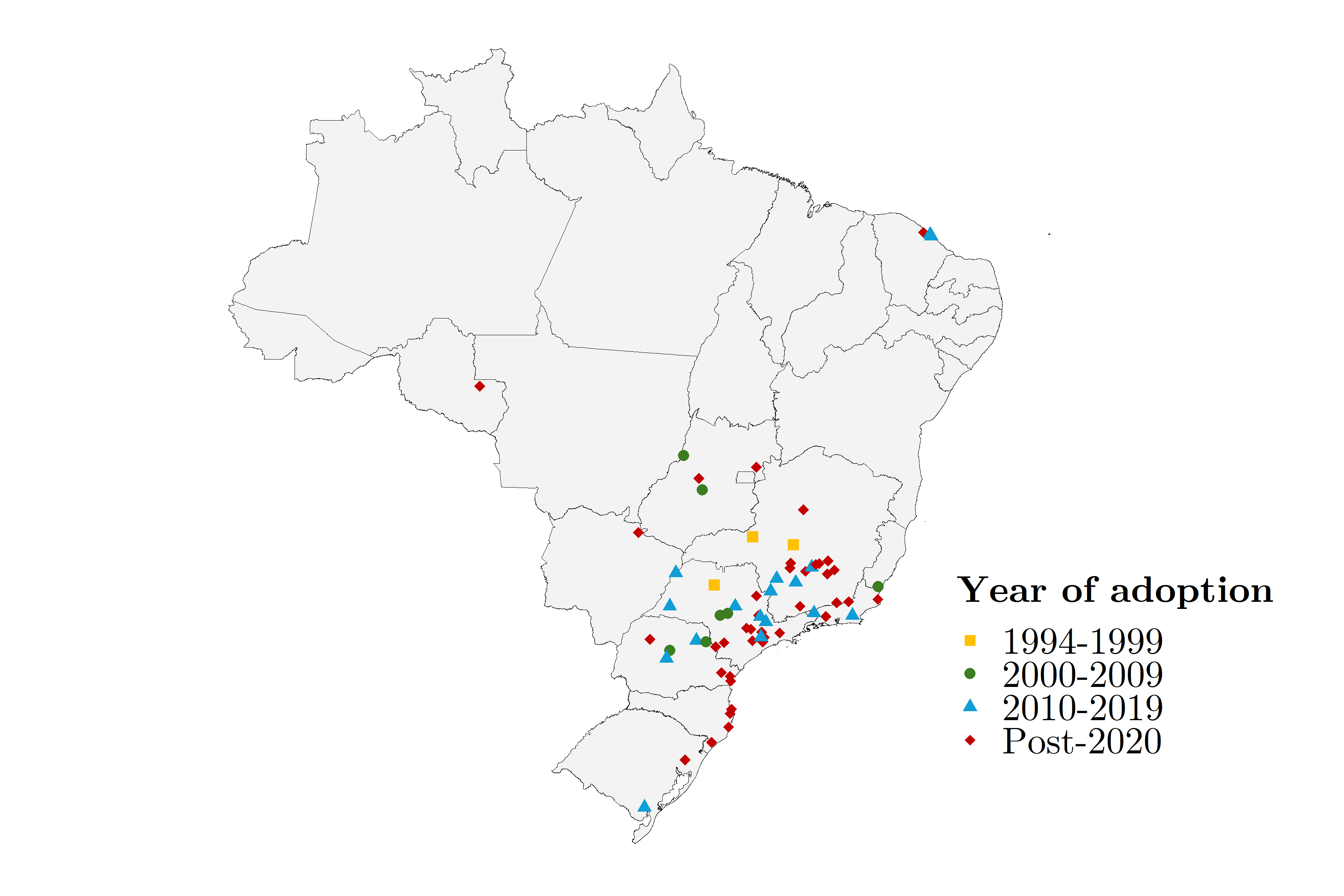}
\end{center}
\footnotesize{\textit{Notes:} This figure displays the location of all municipalities in Brazil that adopted the universal free public transport policy by year of adoption. Yellow squares represent units that were first treated between 1994 and 1999, green dots represent units that were treated between 2000 and 2009, blue triangles indicate units that were first treated between 2010 and 2019, and red losangles are for the units that were treated after 2020. The internal boundaries are the states of Brazil.}
\end{figure}

\begin{figure}[htbp]
\begin{center}
\caption{Number of Municipalities Adopting Universal Free Public Transport}
\label{fig:graph_policy_adoption}
\includegraphics[width=0.70\linewidth]{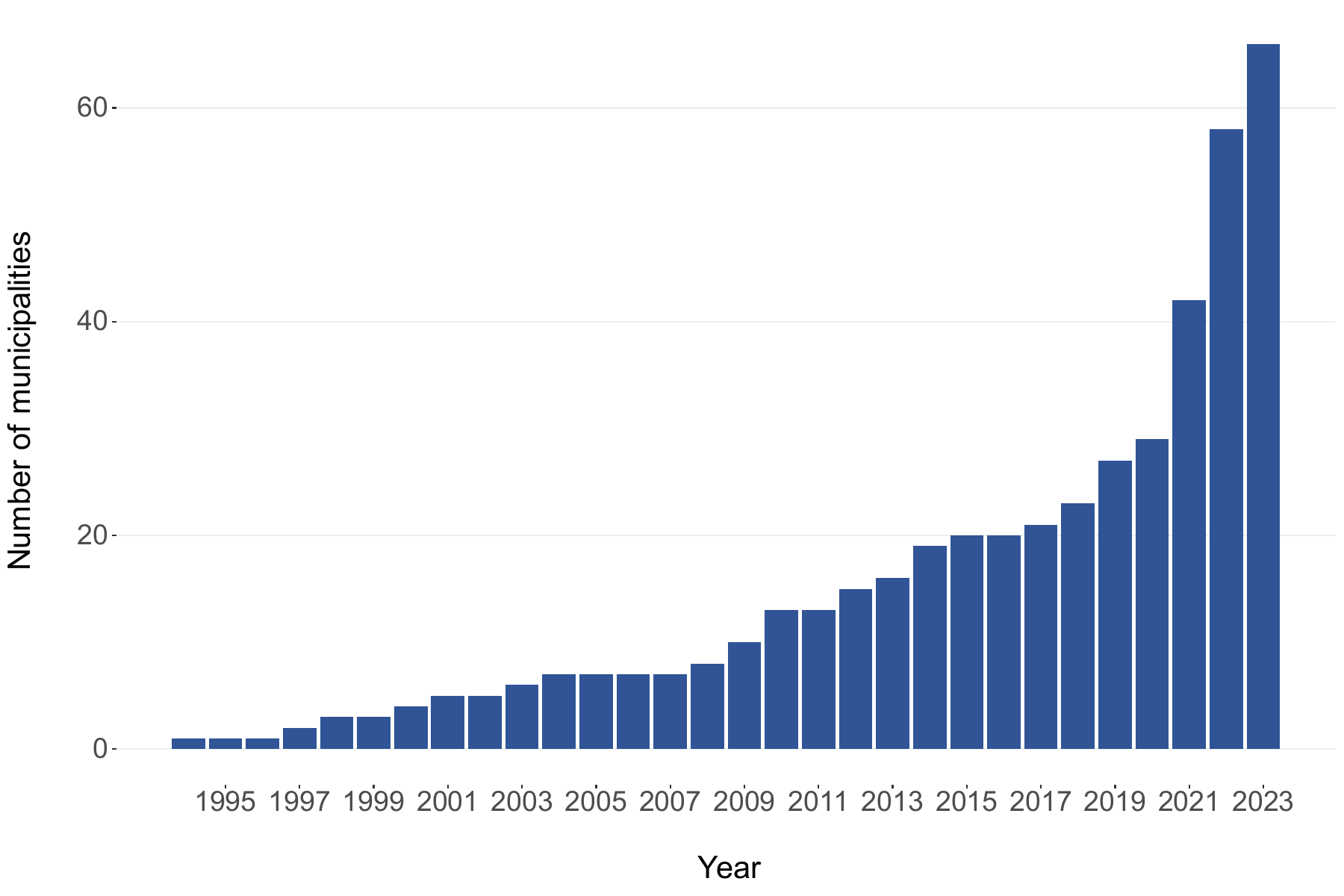}
\end{center}
\footnotesize{\textit{Notes:} This figure displays the cumulative number of municipalities adopting universal free public transport in Brazil. The first treated unit in our sample adopted the policy in 1994, while the last adopted it in 2023.}
\end{figure}

Figure \ref{fig:map_policy_adoption} illustrates the geographical distribution of municipalities implementing universal free public transport. While municipalities in all regions adopt such policies, localities tend to concentrate in the most populous South and Southeast regions. In addition, Figure \ref{fig:graph_policy_adoption} depicts the adoption of the universal free public transport policy over time. The first case occurred in 1994 in the municipality of Monte Carmelo, Minas Gerais, with subsequent adoptions up to recent years. This graph underscores the rollout of the policy adoption, a pattern we explore in our empirical strategy.

% \begin{figure}[htbp]
% \begin{center}
% \caption{Location of Municipalities Adopting Universal Free Public Transport}
% \label{fig:map_policy_adoption}
% \includegraphics[width=0.85\linewidth]{Figs_pdf/mapa_adocao_tarifa_zero.pdf}
% \end{center}
% \footnotesize{\textit{Notes:} This figure displays the location of all municipalities in Brazil that adopted the universal free public transport policy by year of adoption. Yellow squares represent units that were first treated between 1994 and 1999, green dots represent units that were treated between 2000 and 2009, blue triangles indicate units that were first treated between 2010 and 2019, and red losangles are for the units that were treated after 2020. The internal boundaries are the states of Brazil.}
% \end{figure}

% \begin{figure}[htbp]
% \begin{center}
% \caption{Number of Municipalities Adopting Universal Free Public Transport}
% \label{fig:graph_policy_adoption}
% \includegraphics[width=0.70\linewidth]{Files/grafico_adocao_tarifa_zero.png}
% \end{center}
% \footnotesize{\textit{Notes:} This figure displays the cumulative number of municipalities adopting universal free public transport in Brazil. The first treated unit in our sample adopted the policy in 1994, while the last adopted it in 2023.}
% \end{figure}

%%%%%%%%%%%%%%%%%%%%%%%%%%%%%%%%%%%%%%%%
\subsection{Transport, Employment and Emissions}\label{sec:theoretical-mechanisms}
%%%%%%%%%%%%%%%%%%%%%%%%%%%%%%%%%%%%%%%%

In this section, we theoretically discuss how fare-free public transport policies may affect employment and GHG emissions. To test the effects on employment, we hypothesize that fare-free public transport policy can influence labor market outcomes via labor supply, labor demand, and the matching between firms and workers.

After the policy, job search becomes less costly. By promoting greater mobility, the labor search range amplifies, as free public transport allows job-seekers to reach different parts of the city and consider job opportunities in a broader area. Labor search can also be affected by the frequency with which workers search for job positions. Consequently, we would observe an increase in the labor supply of workers directly affected by the policy. Notice, however, that this potential increase in labor supply will translate into higher employment levels---and not only the displacement of existing employed workers---if the policy also affects labor demand or matching. 

On the labor demand side, the savings and facilitated commuting generated by free-fare public transport can positively impact consumption and economic activity. In addition, firms may benefit from improved market access \citep{Tsivanidis2024}. As a result, this policy may influence labor demand via the expansion of existing firms or the creation of new firms, possibly affecting the composition of economic activity.

There is also a potential effect via matching since the policy reduces spatial frictions. For example, workers searching more frequently and reaching a larger area may lead to better matches between their skills and available positions \citep{Agrawal2024}.

In sum, the policy may not mechanically increase employment, and the overall effects depend on what happens to labor supply, labor demand, and matching.

Additionally, free public transport may impact GHG emissions via three channels \citep{Cherniwchan2017}: (i) behavioral changes leading to transport mode change, (ii) growth of economic activity (scale effect), and (iii) composition of economic activity (composition effect).

Behavioral changes regarding transport modes can happen because of cost-saving and environment-related concerns. By making it accessible, more people are likely to use public transport and substitute away from private modes such as cars and motorcycles \citep{dunkerley2018bus}. The fare-free policy may also promote environmental-related behavioral changes: the policy may make the importance of protecting the environment more salient, leading to behavioral changes toward greater use of public transport. These behavioral changes would lead to lower GHG emissions. Another behavioral change from the free policy is a tendency to generate over-consumption of bus rides, given the zero price, which may increase emissions. 

In addition, changes in emissions can be due to the overall effect on economic activity. If the policy stimulates the local economy, one will find an increase in emissions. Finally, the policy can also affect the employment composition, which influences emissions, as different sectors of economic activity have distinct emission intensities. For instance, one worker in livestock is associated with several times more emissions than one worker in services. Therefore, changes in emissions can also happen due to changes in the composition of economic activity measured by compositional changes in the workforce.

As a result of all three channels, the potential mechanisms indicate that the effects of the policy on emissions are conceptually ambiguous.

Lastly, the magnitudes of the effects on employment and emissions will reveal whether there is a decoupling of economic activity---measured in this paper as the variation in employment---from emissions. Decoupling can take two forms: absolute and relative. Absolute decoupling occurs when an increase in economic activity is accompanied by a non-increase in emissions. Relative decoupling means that economic activity and emissions increase, but the increase in economic activity is greater than that of emissions.

%%%%%%%%%%%%%%%%%%%%%%%%%%%%%%%%%%%%%%%%
\section{Data and Sample}\label{sec:data} 
%%%%%%%%%%%%%%%%%%%%%%%%%%%%%%%%%%%%%%%%

We work with several publicly available datasets to build a panel at the municipality-year level. Our period of analysis is 1990--2022.

\medskip
\noindent \textbf{Employment outcome}. Yearly employment data come from RAIS (``Relação Anual de Informações Sociais") dataset. RAIS is a matched employee-employer dataset from Brazil's Ministry of Labor, providing contract-level data on the universe of \textit{formal} workers and firms in Brazil. In this paper, we work with the publicly available RAIS data and aggregate all the information at the municipality-year level.  Therefore, our employment outcome corresponds to the number of formal workers employed in a municipality each year. RAIS also allows us to identify the economic sector, which we use to calculate sectoral employment at the municipality level.\footnote{We use the National Classification of Economic Activities (\textit{Classificação Nacional de Atividades Econômicas}, or CNAE) and restricted the sample to the year 1994 onwards in the sectoral employment analyses to have a consistent definition over time using RAIS. See Appendix Table \ref{tab:tab:sector_aggregation} for more detail on sectorial employment aggregation.} In addition, we use the information on average wages from RAIS.

\medskip
\noindent \textbf{Greenhouse gas emissions}. GHG emission data come from the \textit{Sistema de Estimativas de Emissões e Remoções de Gases de Efeito Estufa} (SEEG), detailed in \cite{seeg2018}. Using guidelines from the Intergovernmental Panel on Climate Change (IPCC), SEEG's methodology entails using data on economic activity and emission factors to measure emissions. Estimates are available for all municipalities combining satellite and field-collected data. Our emission outcome corresponds to yearly total emissions for each municipality. As SEEG data have a standardized methodology from 1990 onwards, we collect data (version released in February 2024) from years 1990 to 2022, using AR6 conversion for the Global Warming Potential of GHG emissions. Greenhouse gases include carbon dioxide (CO2), methane (CH4), nitrous oxide (N2O), and other gases (e.g., perfluorocarbons, hydrofluorocarbons, sulfur hexafluoride, and nitrogen trifluoride). We transform them into an emission metric known as carbon dioxide equivalent (CO$_2$-eq): a common measure that transforms the amounts of other GHG gases to the equivalent amount of CO$_2$ based on their Global Warming Potential. For example, the three main greenhouse gases---CO$_2$, CH$_4$, and N$_2$O---receive weights of 1 (by construction), 28, and 256 when transformed into CO$_2$-eq.

\medskip
\noindent \textbf{Free-fare transport policy}. Data on the year in which each municipally adopted the free-fare transport policy come from the National Association of Public Transport Companies (NTU---\textit{Associação Nacional das Empresas de Transportes Urbanos}). The first municipality in our sample adopted the policy in 1994 and the last one in 2022. 

\medskip
\noindent \textbf{Additional data}. Total population, population in urban areas, per capita income, and average years of schooling stem from the Brazilian Institute of Geography and Statistics (IBGE) Population Census of 1991. These variables are used as covariates in the regression analysis. We use three additional datasets to assess mechanisms. The first contains sales data, in liters, of ethanol and gasoline in each municipality, provided by the National Agency of Petroleum, Natural Gas, and Biofuels (\textit{Agência Nacional do Petróleo, Gás Natural e Biocombustíveis do Brasil}). The second dataset comprises the stock of automobiles in each municipality, provided by the Transport Ministry's National Secretary of Traffic (\textit{Secretaria Nacional de Trânsito}), whose available period is 2002--2022. The third is the publicly available CNPJ (``Cadastro Nacional da Pessoa Jurídica"), a dataset administered by Brazil's Internal Revenue Service. The CNPJ registry has information on the universe of formal firms regarding opening date, and we use these data to create the number of new firms opened by year.

Besides, to perform the cost-benefit analysis, we use microdata from the Brazilian Household Budget Survey (\textit{Pesquisa de Orçamentos Familiares}, or POF) in 2017-2018. The POF survey is conducted by IBGE and details individual expenses and income. In addition, the inflation index IPCA (\textit{Índice de Preços ao Consumidor Amplo}) provided by IBGE and the nominal exchange rate provided by the Brazilian Central Bank are also utilized in our calculations. 

\medskip
\noindent \textbf{Spatial unit of analysis}. In our analysis period, municipalities in Brazil went through a process of detachments. In the year 1990, there were 4,491 municipalities, while in 2022, there were 5,571 units. To deal with the detachments, we use the concept of a Minimum Comparable Area (MCA), which consists of sets of municipalities whose borders were constant over the study period. We thus aggregate municipalities to MCAs. To calculate our outcome variables at the MCA level, we proceed as follows. GHG emissions of an MCA are calculated by summing the emissions from the municipalities that compose it. We perform a similar procedure to calculate the total employment, sectoral employment, fuel sales, and stock of vehicles at the MCA level. To create the treatment indicator variable at the MCA level, we consider a location as treated if any of its municipalities adopt universal free public transport. For ease of exposition, we use the terms municipalities, localities, and MCAs interchangeably henceforth in the paper.

\medskip 
\noindent\textbf{Sample Selection}. Our ``treated” group consists of 56 localities that implemented the universal free-fare transport policy. The treated group has several medium-sized municipalities with population sizes between 100,000 and 500,000 inhabitants. To construct a comparison group more comparable to the treated units, we limit to localities with (i) a population size no greater than the most-populated locality of the treated group and (ii) located in states that have at least one locality that implemented universal free-fare transport policy. We further exclude from the comparison group those localities that implement any form of partially free public transport. We end up with 2,731 localities in the comparison group.\footnote{The outcomes are in logs, so in some outcomes, the number of units decreases---e.g., some municipalities do not have formal workers in one specific economic sector. Appendix Table \ref{tab:sample_selection} shows how the number of units in the comparison group varies according to each step of the sample selection procedure.}

Appendix Table \ref{tab:descriptive_statistics} presents descriptive statistics for each outcome. On average, treated units have more jobs, exhibit higher emissions levels, are more urban, and have more cars and fuel consumption. While these data provide some guidance on levels, we perform a more formal comparison of treated and comparison groups in Section \ref{sec:results}. 

%%%%%%%%%%%%%%%%%%%%%%%%%%%%%%%%%%%%%%%%
\section{Empirical Strategy}\label{sec:empirical_strategy}
%%%%%%%%%%%%%%%%%%%%%%%%%%%%%%%%%%%%%%%%

In this section, we explain how we identify and estimate the effects of free public transport on economic and environmental outcomes. In our empirical application, municipalities choose to adopt free public transport at different points in time and never cease to implement this policy. Consequently, treatment is irreversible and staggered.

For this reason, we adopt a staggered difference-in-differences strategy to identify our target parameters. Although there are many alternatives within this method, we adopt the tools proposed by \cite{callaway2021difference} due to its flexibility with respect to different heterogeneity sources.

In this section, we explain how their method fits our empirical context. We start by formally describing our empirical setting and defining our target parameters (Section \ref{Ssetting}). We then discuss the assumptions that ensure identification of these parameters (Section \ref{Sidentification}). Lastly, we explain our chosen estimation procedure (Section \ref{Sestimation}).

%%%%%%%%%%%%%%%%%%%%%%%%%%%%%%%%%%%%%%%%
\subsection{Setting and Target Parameters}\label{Ssetting}
%%%%%%%%%%%%%%%%%%%%%%%%%%%%%%%%%%%%%%%%
We start by explaining the random variables in our empirical setting. Since treatment is irreversible and staggered, each municipality is part of a group $G$, whose value denotes the first year with free public transport in the municipality. Every municipality also has a vector of potential outcomes (e.g., employment and emissions) that depend on the adoption year and the calendar year. Let $Y_{t}\left(0\right)$ denote the untreated potential outcome at time $t \in \left\lbrace 1, \ldots, T \right\rbrace$ if the municipality does not adopt free public transport during our sample period. Moreover, let $Y_{t}\left(g\right)$ denote the potential outcome at time $t \in \left\lbrace 1, \ldots, T \right\rbrace$ if the municipality starts adopting free public transit in year $g \in \left\lbrace 2, \ldots, T \right\rbrace$.

Now, we explain our target parameters and the types of heterogeneity they capture. Define the group-time average treatment effect: $$ATT\left(g, t\right) = \mathbb{E}\left[\left. Y_{t}\left(g\right) - Y_{t}\left(0\right) \right\vert G = g \right]$$ for any $\left(g,t\right) \in \left\lbrace 2, \ldots, T \right\rbrace \times \left\lbrace 1, \ldots, T \right\rbrace$. This object captures the average treatment effect at time period $t$ for municipalities that are members of a particular group $g$. We aggregate the group-time average treatment effects into three target parameters.

First, we focus on heterogeneity due to the length of exposure to the treatment. To do so, we define $e = t - g$ as the time elapsed since adoption of free public transport and aggregate group-time average treatment effects measured $e$ periods after treatment adoption for all treated municipalities. Formally, we define
\begin{equation}
\label{eq:treatment_effect_length_exposure}
    \theta_{es}(e) := \sum_{g \in \left\lbrace 2, \ldots, T \right\rbrace} \mathbf{1}\{g + e \leq T\}\cdot \mathbb{P}\left[\left. G = g \right\vert G + e \leq T \right] \cdot ATT(g, g + e)
\end{equation}
for each $e \in \left\lbrace 0, \ldots, T - 2\right\rbrace$. This target parameter captures the average effect of adopting free public transport $e$ years after adoption across all municipalities that are ever observed to have taken the treatment for exactly $e$ periods.

Second, we focus on heterogeneity due to adoption year. To do so, we aggregate group-time average treatment effects over calendar years for each adoption year. Formally, we define
\begin{equation}
\label{eq:treatment_effect_group}
    \theta_{sel}(g) := \frac{1}{T - g + 1} \cdot \sum_{t=g}^{T} ATT(g,t)
\end{equation}
for each $g \in \left\lbrace 2, \ldots, T \right\rbrace$. This target parameter captures the average treatment effect for municipalities adopting free public transport for the first time in year $g$, across all their post-treatment years.

Lastly, we focus on a general-purpose summary of the group-time average treatment effects. To do so, we define
\begin{equation}
\label{eq:overall_treatment_effect}
    \theta_{sel}^{O} := \sum_{g \in \left\lbrace 2, \ldots, T \right\rbrace} \theta_{sel}(g) \cdot \mathbb{P}\left[\left. G = g \right\vert G \leq T \right].
\end{equation}
This target parameter captures the average effect of adopting free public transport across all municipalities that ever took the treatment.

%%%%%%%%%%%%%%%%%%%%%%%%%%%%%%%%%%%%%%%%
\subsection{Identifying Assumptions}\label{Sidentification}
%%%%%%%%%%%%%%%%%%%%%%%%%%%%%%%%%%%%%%%%
Besides the assumptions of irreversibility and overlap, we impose two relevant identifying assumptions: ``no anticipation'' and ``unconditional parallel trends for a never-treated group''.

The ``no anticipation'' assumption imposes that, before treatment adoption, the potential outcome of municipalities that eventually adopt free public transport is equal to the untreated potential outcome. Although free public transport policies involve lengthy discussions in the City Council before adoption, this assumption is plausible in our empirical setting. Employment locations and greenhouse gas emissions are choices that are ultimately made by individual citizens instead of mayors or city councilors since these individual citizens are the ones commuting to work and paying for the transit fares, fuel, or car maintenance. When making decisions about their employment location or transport choices, they are likely to consider the current prices instead of possibly free buses in the next year. Consequently, the ``no anticipation'' assumption is simply imposing that, in year $t$, individuals make choices based on year $t$ prices. Moreover, in Section \ref{sec:results}, we test the null hypothesis of ``no anticipation'' and do not reject it.

The ``unconditional parallel trends for a never-treated group'' assumption imposes that there exists a never-treated group whose outcomes evolve similarly to the untreated counterfactual outcomes of the treated municipalities. As explained in Section \ref{sec:data}, we take several steps to construct a never-treated group that is similar to the treated municipalities. Moreover, in Section \ref{sec:results}, we test for parallel pre-trends and we do not reject the null of no parallel pre-trends.

%%%%%%%%%%%%%%%%%%%%%%%%%%%%%%%%%%%%%%%%
\subsection{Estimation and Inference}\label{Sestimation}
%%%%%%%%%%%%%%%%%%%%%%%%%%%%%%%%%%%%%%%%

To estimate the target parameters described in Section \ref{Ssetting}, we use the doubly-robust estimator proposed by \cite{callaway2021difference}. The doubly-robust approach combines the outcome regression method and the inverse probability weighting method and, thus, requires modeling both the outcome expectation and the propensity score. However, for consistency, it only requires either one of those objects to be correctly specified. Consequently, the doubly-robust estimator presents interesting robustness properties against model misspecification when compared to more traditional estimation procedures.

For inference, we cluster standard errors at the municipality level. To construct point-wise confidence bands around our target parameters, we use the multiplier bootstrapped procedure.

%%%%%%%%%%%%%%%%%%%%%%%%%%%%%%%%%%%%%%%%
\section{Results}\label{sec:results}
%%%%%%%%%%%%%%%%%%%%%%%%%%%%%%%%%%%%%%%%
In this section, we present two sets of average effects of adopting free public transport for two outcome variables (employment and greenhouse gas emissions). Section \ref{sec:results-summary} shows the average effect across all municipalities that ever took the treatment (Equation \eqref{eq:overall_treatment_effect}) while Section \ref{sec:results-event-study} shows the average effects for each possible length of exposure to the treatment (Equation \eqref{eq:treatment_effect_length_exposure}).\footnote{In Appendix \ref{sec:results-group}, we show the average effects for each adoption-year group of municipalities (Equation \eqref{eq:treatment_effect_group}).} Moreover, in Section \ref{sec:robustness}, we present three sets of robustness checks.

All our results use the natural logarithms of the outcome variables. Consequently, we interpret our estimates as percentage variations relative to the scenario of no treatment.

%%%%%%%%%%%%%%%%%%%%%%%%%%%%%%%%%%%%%%%%
\subsection{Average Effect of Adopting Free Public Transport Across All Treated Municipalities}\label{sec:results-summary}
%%%%%%%%%%%%%%%%%%%%%%%%%%%%%%%%%%%%%%%%
Our main results focus on a general-purpose summary of all the group-time average treatment effects. This target parameter (Equation \eqref{eq:overall_treatment_effect}) captures the average effect of adopting free public transport across all municipalities that ever took treatment.

Table \ref{tab:overall_att_estimates_seeg_ghg_employment_pandemic} presents the estimates of this target parameter for two outcome variables: the natural logarithm of the formally employed individuals in each municipality (Column (1)) and the natural logarithm of CO\textsubscript{2}-equivalent emissions in each municipality (Column (2)). These results are based on a doubly-robust estimator and standard errors are clustered at the municipality level.

% \begin{table}[htbp]
% \begin{center}
% \caption{\label{tab:overall_att_estimates_seeg_ghg_employment_pandemic} Average Effect of Adopting Free Public Transit Across All Treated Municipalities}
% \begin{tabular}{lcc}
% \hline \hline
% & \multicolumn{2}{c}{Outcome Variable} \\ \hline
%  & Employment & GHG emissions \\
%  & (1) & (2) \\
% \hline
% ATT (Summary) & $0.038^{***}$ & $-0.043^{*}$ \\
%  & (0.011) & (0.022) \\
% \hline
%  Units & 2361 & 2308 \\
% Treated units & 41 & 56 \\
% Groups & 19 & 20 \\
% \hline
% \end{tabular}
% \end{center}
% \footnotesize{\textit{Notes:} This table presents the estimates of the average effect of adopting free public transport across all municipalities that ever took treatment (Equation \eqref{eq:overall_treatment_effect}). The outcome variables are the natural logarithm of the formally employed individuals in each municipality (Column (1)) and the natural logarithm of CO\textsubscript{2}-equivalent emissions in each municipality (Column (2)). These results are based on the doubly-robust estimator proposed by \cite{callaway2021difference}. Standard errors are reported in parenthesis and are clustered at the municipality level. At the bottom, we also report the total number of municipalities in our samples, the number of treated municipalities and the number of adoption-year groups. Significance levels are denoted as follows: $^{***}p<0.01$; $^{**}p<0.05$; $^{*}p<0.1$.}
% \end{table}

\begin{table}[htbp]
\begin{center}
\caption{\label{tab:overall_att_estimates_seeg_ghg_employment_pandemic} Average Effect of Adopting Free Public Transit Across All Treated Municipalities}
\begin{tabular}{lcc}
\toprule \toprule
 & Employment & GHG emissions\\
\midrule
Treatment effect & $0.032^{***}$ & $-0.041^{*}$\\
 & (0.011) & (0.021)\\
\midrule
Treated units & 57 & 57\\
Groups & 20 & 20\\
Units & 2,361 & 2,371\\
\bottomrule \bottomrule
\end{tabular}
\end{center}
\footnotesize{\textit{Notes:} This table presents the estimates of the average effect of adopting free public transport across all municipalities that ever took treatment (Equation \eqref{eq:overall_treatment_effect}). The outcome variables are the natural logarithm of the formally employed individuals in each municipality (Column (1)) and the natural logarithm of CO\textsubscript{2}-equivalent emissions in each municipality (Column (2)). These results are based on the doubly-robust estimator proposed by \cite{callaway2021difference}. Standard errors are reported in parentheses and are clustered at the municipality level. At the bottom, we also report the total number of municipalities in our samples, the number of treated municipalities and the number of adoption-year groups. Significance levels are denoted as follows: $^{***}p<0.01$; $^{**}p<0.05$; $^{*}p<0.1$.}
\end{table}

First, note that free public transport significantly increases the formal employment level by approximately 3.2\%. These estimates may be considered large when compared to related policies previously discussed in the literature. For instance, \cite{tyndall2021local} analyzes a light rail transit expansion and finds that it reduced aggregate metropolitan employment.

Second, observe that free public transport significantly decreases greenhouse gas emissions by approximately 4.1\%. Once more, our estimates may be considered large when compared to related policies previously discussed in the literature. For example, \cite{lin2021impact} analyzes the impact of constructing high-speed railroads in China and finds a negative effect of only 1.33\% on emissions. However, we highlight that our outcome variable considers all economic sectors while their outcome variable considers only transport sector emissions.

Combining these two results, we conclude that free public transport generates absolute decoupling by boosting economic activity while decreasing greenhouse gas emissions. In other words, these policies reduce emissions intensity (defined here as emissions per formal worker).

%%%%%%%%%%%%%%%%%%%%%%%%%%%%%%%%%%%%%%%%
\subsection{Average Effect for each Length of Exposure to the Treatment}\label{sec:results-event-study}
%%%%%%%%%%%%%%%%%%%%%%%%%%%%%%%%%%%%%%%%

We now show the average effects for each possible length of exposure to the treatment (Equation \eqref{eq:treatment_effect_length_exposure}). These target parameters are useful for two reasons: they allow us to (i) test the null that time trends between ``eventually treated units'' and ``never treated units'' are parallel before free public transit is adopted, and (ii) to discuss whether average treatment effects are increasing, decreasing, or constant as the time since treatment adoption increases.

\begin{figure}[htbp]
\begin{center}
\caption{Average Effects for each Length of Exposure to the Treatment}
\label{fig:gwp_100_ar5_ghg_vinculos_ativos_dynamic_pandemia}
\begin{subfigure}{0.8\linewidth}
\caption{Outcome Variable: Employment}
\includegraphics[width=1\linewidth]{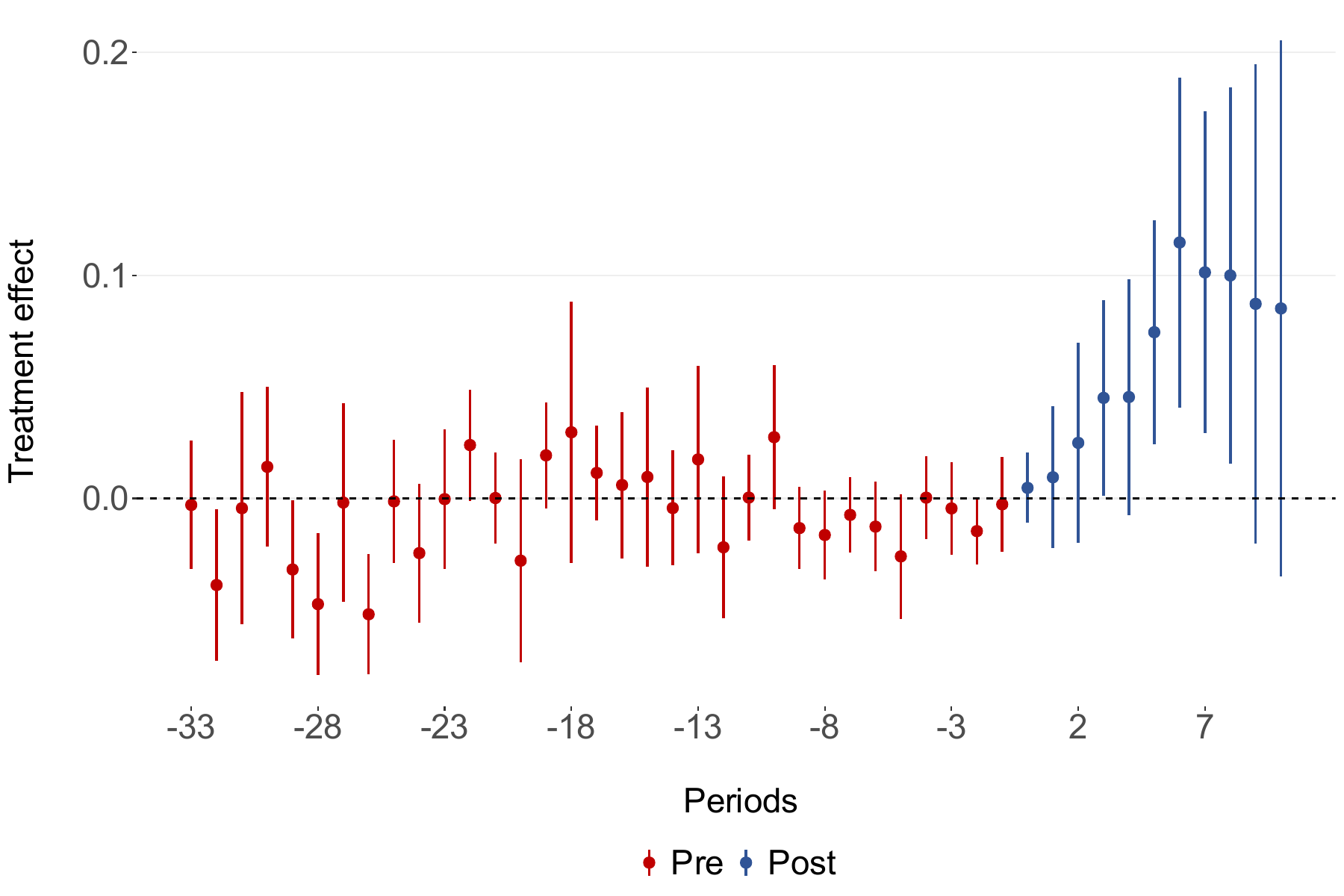}
\label{fig:vinculos_ativos_dynamic_pandemia}
\end{subfigure} \\
\begin{subfigure}{0.8\linewidth}
\caption{Outcome Variable: Greenhouse Gas Emissions}
\includegraphics[width=1\linewidth]{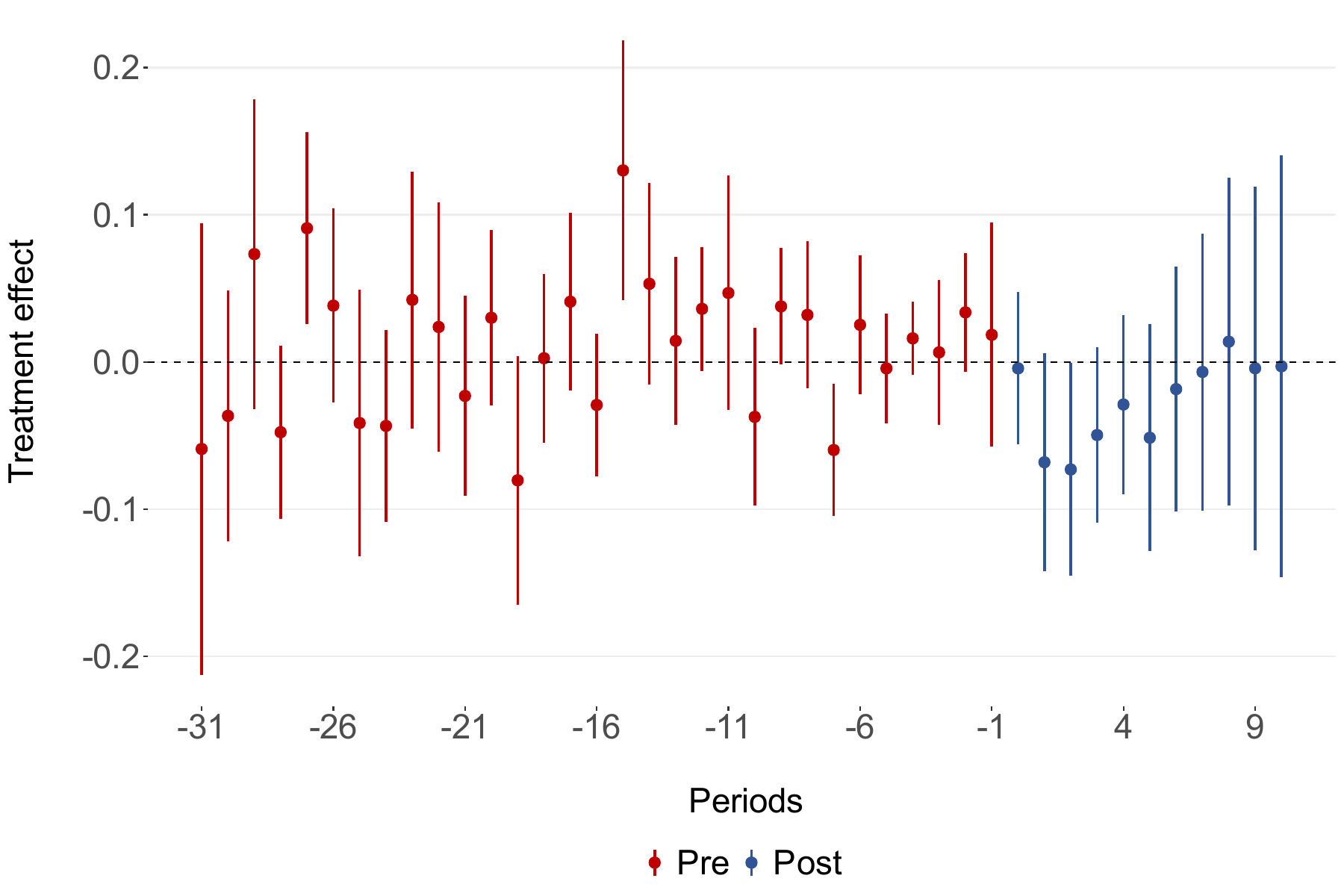}
\label{fig:gwp_100_ar5_ghg_dynamic_pandemia}
\end{subfigure}
\end{center}
\footnotesize{\textit{Notes:} Figure \ref{fig:gwp_100_ar5_ghg_vinculos_ativos_dynamic_pandemia} presents estimates of the average effect of adopting free public transit $e$ years after adoption across all municipalities that are ever observed to have taken the treatment for exactly $e$ periods (Equation \eqref{eq:treatment_effect_length_exposure}). The outcome variables are the natural logarithm of the formally employed individuals in each municipality (Figure \ref{fig:vinculos_ativos_dynamic_pandemia}) and the natural logarithm of CO\textsubscript{2}-equivalent emissions in each municipality (Figure \ref{fig:gwp_100_ar5_ghg_dynamic_pandemia}). Vertical lines represent point-wise 90\%-confidence intervals based on standard errors clusterized at the municipality level. These results are based on the doubly-robust estimator proposed by \cite{callaway2021difference} using a varying base period for the pre-treatment placebo estimates (in red). Post-treatment estimates are reported in blue.}
\end{figure}

Figure \ref{fig:gwp_100_ar5_ghg_vinculos_ativos_dynamic_pandemia} presents estimates of the average effect of adopting free public transit $e$ years after adoption across all municipalities that are ever observed to have taken the treatment for exactly $e$ periods (Equation \eqref{eq:treatment_effect_length_exposure}). Again, the outcome variables are the natural logarithm of the formally employed individuals (Figure \ref{fig:vinculos_ativos_dynamic_pandemia}) and the natural logarithm of CO\textsubscript{2}-equivalent emissions (Figure \ref{fig:gwp_100_ar5_ghg_dynamic_pandemia}). Vertical lines represent point-wise 90\%-confidence intervals based on standard errors clustered at the municipality level. While post-treatment effects are reported in blue, pre-treatment placebo estimates are reported in red.

These results are based on a doubly-robust estimator. Differently from most event-study papers, we use a varying base period for the pre-treatment placebo estimates (in red) instead of using a universal base period.\footnote{As \cite{callaway2021package} explain, either type of base period can be written as a linear combination of the other. Consequently, they are just alternative ways of reporting the same information even though they have different interpretations.} Consequently, the red dots capture pseudo-ATTs, i.e., our pre-treatment estimates represent what we would estimate as the immediate policy effect if the policy occurred in that period instead of period $0$.\footnote{In other words, we estimate the immediate placebo effect if the policy was artificially assigned to start in a period before the actual treatment.} We choose to report pre-treatment results with a varying base period because it is more intuitive to use them to test the ``no anticipation assumption''.

We start by discussing the results related to employment levels and, then, we discuss the results related to greenhouse gas emissions.

First, when we use employment as our outcome variable (Figure \ref{fig:vinculos_ativos_dynamic_pandemia}), we find that the pre-treatment placebo estimate just before the treatment (red dot in period -1) is statistically null. This result suggests that there are no anticipation effects of adopting free public transport on employment. Moreover, almost all pre-treatment placebo estimates (red dots) in Figure \ref{fig:vinculos_ativos_dynamic_pandemia} are statistically null and approximately half of them are positive, while the other half is negative. These results suggest that the parallel trends assumption is plausible in the pre-treatment period when we use employment as our outcome variable.

Second, observe that post-adoption treatment effect estimates (blue dots) in Figure \ref{fig:vinculos_ativos_dynamic_pandemia} are large and statistically significant for some periods when we use employment as our outcome variable. Moreover, these point estimates suggest that average treatment effects on employment might be increasing as a function of the length of exposure to the treatment. Our point-estimates increase in the first seven years and then become stable around an estimated effect of 10\%.

Third, when we use greenhouse gas emissions as our outcome variable (Figure \ref{fig:gwp_100_ar5_ghg_dynamic_pandemia}), we find that the pre-treatment placebo estimate just before the treatment (red dot in period -1) is statistically null. This result suggests that there are no anticipation effects of adopting free public transport on emissions. Moreover, almost all pre-treatment placebo estimates (red dots) in Figure \ref{fig:gwp_100_ar5_ghg_dynamic_pandemia} are statistically null, and approximately two-thirds of them are positive, while the other third is negative. These results suggest that the parallel trends assumption is plausible in the pre-treatment period when we use emissions as our outcome variable.

Lastly, observe that post-adoption treatment effect estimates (blue dots) in Figure \ref{fig:gwp_100_ar5_ghg_dynamic_pandemia} are negative (but not precisely estimated) or small when we use emissions as our outcome variable. Moreover, these point estimates suggest that adopting free public transit may have negative short-run effects on emissions with zero long-run effects. 

%%%%%%%%%%%%%%%%%%%%%%%%%%%%%%%%%%%%%%%%
\subsection{Robustness Checks}\label{sec:robustness}
%%%%%%%%%%%%%%%%%%%%%%%%%%%%%%%%%%%%%%%%
In this section, we present three sets of robustness checks: exclusion of the years affected by the COVID-19 Pandemic, addition of control variables in our Staggered Differences-in-Differences analyses, and removal of the last year of the employment data due to changes in the data collection methodology.\footnote{In Appendix \ref{sec:results-group}, we also discuss one extra heterogeneity source: the average effects for each adoption-year group of municipalities (Equation \eqref{eq:treatment_effect_group}).}

In the main results, our sample includes the years during the COVID-19 Pandemic (2020, 2021 and 2022). We choose to include them to use the entire sample of municipalities. As a robustness check, Appendix \ref{appendix_covid} presents the estimates of our target parameters when we remove these three years from our sample. These results are similar to those presented in the main text.

We perform the main analysis without control variables. Consequently, our main identification strategy relies on an unconditional parallel trends assumption between our treated and untreated groups. Considering that these treated municipalities seem larger and more developed than untreated municipalities (Table \ref{tab:descriptive_statistics}), we may be concerned that this identification assumption is too strong. For this reason, we also adopt a conditional parallel trends assumption that imposes that the expected value of the untreated potential outcome follows the same trend in the treated and untreated groups after conditioning on a set of covariates. To do so, we re-run our Staggered Differences-Differences analyses controlling for population size, urban population share, per capita income, and average years of schooling of each municipality (using values from the 1991 Population Census interacted with time fixed effects). Appendix Table \ref{tab:overall_att_estimates_with_controls} shows the results when we include these covariates in the estimation of the
general-purpose summary of all the group-time average treatment effects (Equation \eqref{eq:overall_treatment_effect}). Reassuringly, the results are robust to the inclusion of covariates. 

The RAIS dataset changed its collection methodology in 2022. Therefore, we run the employment analysis using data until 2021 to check for robustness. Appendix Table \ref{tab:att_estimates_seeg_ghg_employment_until2021} shows that the results are similar when we exclude the year 2022 from the analysis.

%%%%%%%%%%%%%%%%%%%%%%%%%%%%%%%%%%%%%%%%
\section{Further Analyses: Potential Mechanisms and Cost-Benefit Analysis}\label{sec:further_analyses}
%%%%%%%%%%%%%%%%%%%%%%%%%%%%%%%%%%%%%%%%
In this section, we deepen our analysis of free public transport by discussing the potential mechanisms through which this policy affects the local economy (Section \ref{sec:mechanisms}) and implementing a cost-benefit analysis (Section \ref{sec:cost-benefit}).

Considering our target policy's potential mechanisms, we conjecture that free public transport allows rural job-seekers to amplify their job search, allowing rural workers to find city-oriented formal jobs, possibly increasing economic activity. If this hypothesized transformation is true, it also explains the negative effect on greenhouse gas emissions because agriculture is a higher-emission sector in Brazil, while urban sectors are associated with lower emissions \citep{DaMata2024}.

Related to our target policy's cost and benefits, we estimate (i) the government expenditures due to larger subsidies to the public transport system, (ii) the environmental benefits due to reduced carbon emissions, and (iii) the tax revenue due to more formal jobs. Although the increased subsidies are larger than each type of benefit separately, the combined benefits surpass this policy's costs.

%%%%%%%%%%%%%%%%%%%%%%%%%%%%%%%%%%%%%%%%
\subsection{Potential Mechanisms}\label{sec:mechanisms}
%%%%%%%%%%%%%%%%%%%%%%%%%%%%%%%%%%%%%%%%
In this section, we discuss the potential mechanisms behind the effects of free public transport, focusing on possible explanations for the decoupling of economic activity from greenhouse gas emissions (Section \ref{sec:theoretical-mechanisms}). To do so, we analyze our target policy's effects on (i) the stock of automobiles, (ii) fuel sales, (iii) creation of new firms, and (iv) formal employment by economic sector.

First, we want to understand whether individuals refrain from using private transport because of our target policy. Data restrictions do not allow us to directly analyze the effect of free public transit on the use of public transport, as bus travel data are not measured or available for several municipalities of our sample. However, we are able to assess the effects of the policy on the use of private transport, for which data are available. More precisely, we analyze the effect of free public transport on automobile stock, gasoline sales, and ethanol sales.\footnote{We do not use diesel sales as an additional outcome variable for two reasons. First, private cars in Brazil typically use either gasoline or ethanol. According to the National Association of Producers of Vehicles (\textit{Associação Nacional dos Fabricantes de Veículos Automotores}, or ANFAVEA), in 2022, 85.8\% of the sales of automobiles in Brazil were either fueled by gasoline or ethanol. Second, diesel is also used for manufacturing, implying that its sales would be influenced not only by the switch from cars to buses but also by economic activity.} If individuals opt out of private transport, they will reduce their fuel consumption and may sell their vehicles.

Table \ref{tab:overall_att_estimates_stock_automobiles_fuel_sales_pandemic} presents estimates of our general-purpose summary of all the group-time average treatment effects (Equation \eqref{eq:overall_treatment_effect}) for three outcome variables: the natural logarithm of the stock of automobiles (Column (1)), the natural logarithm of gasoline sales (Column (2)) and the natural logarithm of ethanol sales (Column (3)). These results are based on a doubly-robust estimator, and standard errors are clustered at the municipality level.

% \begin{table}[htbp]
% \begin{center}
% \caption{\label{tab:overall_att_estimates_stock_automobiles_fuel_sales_pandemic}Average Effect of Adopting Free Public Transit Across All Treated Municipalities \\ Outcome Variables: Stock of Automobiles and Fuel Sales}
% \begin{tabular}{lccc}
% \hline \hline
%  & Stock of Automobiles & Gasoline sales & Ethanol sales\\
%  & (1) & (2) & (3) \\
% \hline
% ATT (Summary) & $0.014^{}$ & $0.121^{}$ & $-0.01^{}$\\
%  & (0.026) & (0.101) & (0.008)\\
% \hline
%  Units & 2,162 & 1,860 & 2,364\\
% Treated units & 55 & 52 & 52\\
% Groups & 20 & 19 & 15\\
% \hline
% \end{tabular}
% \end{center}
% \footnotesize{\textit{Notes:} This table presents the estimates of the average effect of adopting free public transport across all municipalities that ever took treatment (Equation \eqref{eq:overall_treatment_effect}). The outcome variables are the natural logarithm of the stock of automobiles (Column (1)), gasoline sales (Column (2)) and ethanol sales (Column (3)). These results are based on the doubly-robust estimator proposed by \cite{callaway2021difference}. Standard errors are reported in parenthesis and are clustered at the municipality level. At the bottom, we also report the total number of municipalities in our samples, the number of treated municipalities and the number of adoption-year groups. Significance levels are denoted as follows: $^{***}p<0.01$; $^{**}p<0.05$; $^{*}p<0.1$.}
% \end{table}

\begin{table}[H]
\begin{center}
\caption{\label{tab:overall_att_estimates_stock_automobiles_fuel_sales_pandemic} Average Effect of Adopting Free Public Transit Across All Treated Municipalities \\ Outcome Variables: Stock of Automobiles, Fuel Sales, and Firm Creation}
\scalebox{0.9}{
\begin{tabular}{lcccc}
\toprule \toprule
 & Stock of automobiles & Gasoline sales & Ethanol sales & Firm creation\\
 & \footnotesize{(1)} & \footnotesize{(2)} & \footnotesize{(3)} & \footnotesize{(4)}\\
\midrule
Treatment effect & $-0.01^{}$ & $0.014^{}$ & $0.121^{}$ & $0.074^{***}$\\
 & (0.008) & (0.027) & (0.10)  & (0.018)\\
\midrule
Treated units & 52 & 55 & 52 & 57\\
Groups & 15 & 20 & 19 & 20\\
Units & 2364 & 2162 & 1860 & 2308\\
\bottomrule \bottomrule
\end{tabular}}
\end{center}
\footnotesize{\textit{Notes:} This table presents the estimates of the average effect of adopting free public transport across all municipalities that ever took treatment (Equation \eqref{eq:overall_treatment_effect}). The outcome variables are the natural logarithm of the stock of automobiles (Column (1)), gasoline sales (Column (2)) and ethanol sales (Column (3)). These results are based on the doubly-robust estimator proposed by \cite{callaway2021difference}. Standard errors are reported in parentheses and are clustered at the municipality level. At the bottom, we also report the total number of municipalities in our samples, the number of treated municipalities and the number of adoption-year groups. Significance levels are denoted as follows: $^{***}p<0.01$; $^{**}p<0.05$; $^{*}p<0.1$.}
\end{table}

We find null effects for all three outcome variables. These results suggest that the decrease in greenhouse gas emissions is not driven by individuals reducing the use of private transport. Notice that while the magnitudes for automobiles and gasoline are closer to zero, the effect on ethanol sales is positive, but not precisely estimated. Appendix Figure \ref{fig:automovel_vendas_combustiveis_dynamic} depicts our estimates of the average effects for each length of exposure for these three outcome variables. Its results suggest that there are no anticipation effects of adopting free public transport and that the parallel trends assumption is plausible in the pre-treatment period for these outcome variables.

Second, we assess the impacts on the creation of new firms. Column (4) of Table \ref{tab:overall_att_estimates_stock_automobiles_fuel_sales_pandemic} presents estimates of our general-purpose summary of all the group-time average treatment effects (Equation \eqref{eq:overall_treatment_effect}) for an outcome variable that corresponds to the natural logarithm of new firms created in each municipality per year. The results indicate an increase of 7\% in the number of new firms. This increase in entrepreneurship and the creation of firms are consistent with a rise in labor demand influencing the increased aggregate employment we observe.

Third, we want to understand the role of sectoral change in response to the policy. To do so, we analyze formal employment levels in six economic sectors: manufacturing, construction, commerce, services, agriculture, and transportation.\footnote{See Appendix Table \ref{tab:tab:sector_aggregation} for the formal definition of each sector.} If free public transport policies enlarge workers' job searching area (and improve employer-employee matches or positively impact labor demand), we may find that city-oriented sectors that benefit more from the expanded search increase formal employment. Moreover, if individuals move from higher-emission to lower-emission sectors, the changing job composition at the local level is consistent with the observed increased formal employment and decreased GHG emissions, as sectors of economic activity have different levels of GHG-emission intensity. 

Table \ref{tab:overall_att_estimates_employment_sector_pandemic} presents estimates of our general-purpose summary of all the group-time average treatment effects (Equation \eqref{eq:overall_treatment_effect}) for six outcome variables: the natural logarithm of the formal employment level in manufacturing (Column (1)), construction (Column (2)), commerce (Column (3)), services (Column (4)), agriculture (Column (5)), and transportation (Column (6)). These results are based on a doubly-robust estimator and standard errors are clustered at the municipality level.

\begin{table}[htb]
\begin{center}
\caption{\label{tab:overall_att_estimates_employment_sector_pandemic}Average Effect of Adopting Free Public Transit Across All Treated Municipalities \\ Outcome Variable: Formal Employment Level by Sector}
\begin{tabular}{lccc}
\toprule \toprule
 & \multicolumn{3}{c}{Panel A} \\
 \cline{2-4}
 & Manufacturing & Construction & Commerce\\
 & \footnotesize{(1)} & \footnotesize{(2)} & \footnotesize{(3)}\\
\midrule
ATT (Summary) & $0.099^{}$ & $0.247^{*}$ & $-0.015^{}$ \\
 & (0.067) & (0.149) & (0.021) \\
\midrule
Treated units & 40 & 31 & 40 \\
Groups & 18 & 14 & 18 \\
Units & 1981 & 976 & 2202 \\
\bottomrule 
\bottomrule
& \multicolumn{3}{c}{Panel B} \\
 \cline{2-4}
 & Services & Agriculture & Transportation\\
 & \footnotesize{(4)} & \footnotesize{(5)} & \footnotesize{(6)} \\
\midrule
ATT (Summary) & $0.027^{}$ & $-0.049^{**}$ & $-0.056^{}$\\
 & (0.019) & (0.021) & (0.046)\\
\hline
Treated units & 40 & 39 & 39\\
Groups & 18 & 18 & 18\\
Units & 2324 & 2161 & 2212\\
\bottomrule \bottomrule
\end{tabular}
\end{center}
\footnotesize{\textit{Notes:} This table presents the estimates of the average effect of adopting free public transit across all municipalities that ever took treatment (Equation \eqref{eq:overall_treatment_effect}). The outcome variables are the natural logarithms of formal employment in six economic sectors: Manufacturing (Column (1)), Construction (Column (2)), Commerce (Column (3)), Services (Column (4)), Agriculture (Column (5)), Transportation (Column (6)). These sectors are formally defined in Table \ref{tab:tab:sector_aggregation}. These results are based on the doubly-robust estimator proposed by \cite{callaway2021difference}. Standard errors are reported in parentheses and are clustered at the municipality level. At the bottom, we also report the total number of municipalities in our samples, the number of treated municipalities and the number of adoption-year groups. Significance levels are denoted as follows: $^{***}p<0.01$; $^{**}p<0.05$; $^{*}p<0.1$.}
\end{table}

We find null effects for four out of six economic sectors. The only significant results are an increase in construction employment and a decrease in agricultural employment. These results suggest that the reduction in greenhouse gas emissions (Section \ref{sec:results-summary}) is driven by a compositional change in economic activity: workers are moving from higher-emission sectors (agriculture) to city-oriented, lower-emission sectors (construction). This sector composition change also explains the absolute decoupling of economic activity and greenhouse gas emissions even without any impact on the use of private transportation.

Appendix Figure \ref{fig:vinculos_ativos_setor_dynamic} shows our estimates of the average effects for each length of exposure. It suggests that there are no anticipation effects and that the parallel trends assumption is plausible in the pre-treatment period for these outcome variables. 

We conjecture that free public transport policies may allow individuals who live in rural areas and work in agriculture to seek employment in urban areas and city-oriented sectors. This movement may increase overall formal employment if individuals previously working in home production or informal agricultural activities find jobs in formal firms in urban areas due to an improved matching technology (for instance, due to a larger job-searching area).

If this hypothesized rural-urban movement is true, it also explains the null effect on the stock of automobiles and fuel sales. In this scenario, individuals who did not use buses or cars in rural areas are the ones finding jobs in construction in urban areas. Since they previously did not use private transport, our target policy would not have an effect on the stock of automobiles or fuel sales.

%%%%%%%%%%%%%%%%%%%%%%%%%%%%%%%%%%%%%%%%
\subsection{Cost-benefit Analysis}\label{sec:cost-benefit}
%%%%%%%%%%%%%%%%%%%%%%%%%%%%%%%%%%%%%%%%
To evaluate the costs and benefits of a free public transport policy, we estimate (i) the government expenditures due to larger subsidies to the public transit system, (ii) the environmental benefits due to reduced carbon emissions, and (iii) the tax revenue due to more formal jobs.

\medskip 
\noindent\textbf{Government Expenditures}. To estimate the government expenditures due to larger subsidies related to the public transport system, we use the urban transport expenses from the POF 2017-2018 microdata and find that the average annual expense for urban transport in Brazil is 123.48 measured in Brazilian Reais of 2021 according to the IPCA inflation index.\footnote{We assigned the value zero to individuals without urban transport expenses.} By multiplying this value by the population of the average treated municipality each year, we find the average annual cost that local governments need to finance to provide free public transit.

\medskip 
\noindent\textbf{Environmental Benefits}. The environmental benefits due to reduced greenhouse gas emissions are measured in monetary value using the social cost of carbon, which captures the economic cost associated with the emission of one additional ton of carbon equivalent. More precisely, we use the following procedure to calculate the environmental benefit of free public transit.

For each treated municipality $i$ and each year $t$ after free public transport adoption, we define the realized natural logarithm of greenhouse gas emissions $(\ln{E_{i,t}})$. We, then, estimate the untreated counterfactual natural logarithm of emissions $(\widehat{\ln{E}}_{i,t}^{0})$ by subtracting the estimated ATT (Table \ref{tab:overall_att_estimates_seeg_ghg_employment_pandemic}) from the realized natural logarithm of emissions, i.e., $$\widehat{\ln{E}}_{i,t}^{0} \coloneqq  \ln{E_{i,t}} - (-0.041).$$ Afterward, we apply the exponential function to both objects to find the realized level and the untreated counterfactual level of emissions, i.e., $$E_{i,t} \coloneqq \exp{\left(\ln{E_{i,t}}\right)} \text{ and } \widehat{E}_{i,t}^{0} \coloneqq \exp{\left(\widehat{\ln{E}}_{i,t}^{0}\right)}.$$

Next, we estimate the average reduction in greenhouse gas emissions in each year due to the adoption of free public transport, i.e., $$\overline{\Delta E_{t}} \coloneqq - \left\lbrace \sum\limits_{i \in \text{Treated Cities in year $t$}} \left(\dfrac{E_{i,t} - \widehat{E}_{i,t}^{0}}{N_{1,t}}\right) \right\rbrace,$$ where $N_{1,t}$ is the number of treated municipalities in year $t$.

Lastly, to compute the monetary value of the environmental benefits due to reduced greenhouse gas emissions $\left(B^{env}\right)$, we use the social cost of carbon to measure $\overline{\Delta E_{t}}$ in Brazilian reais of 2021, i.e.,
\begin{equation}
    \label{eq:carbon_benefit}
    B_{t}^{env} \coloneqq SC_{2020} \cdot \varepsilon_{2020} \cdot \frac{P_{2021}}{P_{2020}} \cdot \overline{\Delta E_{t}},
\end{equation}
where $SC_{2020}$ denotes the social cost of carbon emissions in 2020 based on the average estimate with a 2.5\% discount rate according to \cite{us2021social}, $\varepsilon_{2020}$ is the end-of-period exchange rate between Brazilian reais and American dollars in 2020, and $P_{2021}$ and $P_{2020}$ are the Brazilian price levels in 2021 and 2020 according to the IPCA index.\footnote{We use the value of 76 dollars per metric of ton of CO2 in 2020 for the social cost of carbon \citep{us2021social}.}

\medskip 
\noindent\textbf{Tax Revenue}. We also estimate the fiscal externalities from the policy, i.e., the tax revenue due to the formal jobs created by the free public transport policy. We measure these fiscal benefits in monetary value by combining the estimated effect on formal employment (Table \ref{tab:overall_att_estimates_seeg_ghg_employment_pandemic}) with the average rate of payroll and income taxes.\footnote{We do not take into consideration the treatment effect on the average formal wage in the municipality because we estimated it and found a null effect. We present detailed results for the average treatment effect on formal wages in Appendix Table \ref{tab:salarios} and Appendix Figure \ref{fig:salario}.} More precisely, we use the following procedure to calculate the fiscal benefit of free public transit.

For each treated municipality $i$ and each year $t$ after free public transport adoption, we define the realized natural logarithm of employment $(\ln{J_{i,t}})$. We, then, estimate the untreated counterfactual natural logarithm of employment $(\widehat{\ln{J}}_{i,t}^{0})$ by subtracting the estimated ATT (Table \ref{tab:overall_att_estimates_seeg_ghg_employment_pandemic}) from the realized natural logarithm of employment, i.e., $$\widehat{\ln{J}}_{i,t}^{0} \coloneqq  \ln{J_{i,t}} - 0.032.$$ Afterward, we apply the exponential function to both objects to find the realized level and the untreated counterfactual level of employment, i.e., $$J_{i,t} \coloneqq \exp{\left(\ln{J_{i,t}}\right)} \text{ and } \widehat{J}_{i,t}^{0} \coloneqq \exp{\left(\widehat{\ln{J}}_{i,t}^{0}\right)}.$$

Next, we estimate the average increase in employment in each year due to the adoption of free public transport, i.e., $$\overline{\Delta J_{t}} \coloneqq \sum\limits_{i \in \text{Treated Cities in year $t$}} \left(\dfrac{J_{i,t} - \widehat{J}_{i,t}^{0}}{N_{1,t}}\right),$$ where $N_{1,t}$ is the number of treated municipalities in year $t$.

Furthermore, we estimate the increase in aggregated labor income by multiplying the increase in employment by the average annual wage in real prices, i.e., $$\overline{\Delta LI_{t}} \coloneqq \overline{\Delta J_{t}} \cdot \overline{W_{t}} \cdot 13.33 \cdot \dfrac{P_{2021}}{P_{t}},$$ where $\overline{W_{t}}$ is the average formal monthly wage in the treated municipalities in year $t$, 13.33 is the number of monthly wages paid in a year according to Brazilian Law, and $P_{t}$ is the Brazilian price level in year $t$ according to the IPCA index.\footnote{The publicly available RAIS dataset contains wage information from 1999 onwards. To estimate the average wage between 1994 and 1998, we take the average wage in 1999 and deflate it according to the inflation experienced during those years.}

Lastly, to compute the monetary value of the fiscal benefits $\left(B_{t}^{fiscal}\right)$, we multiply the increased labor income $(\overline{\Delta LI_{t}})$ by the average rate of payroll and income taxes $(\tau)$, i.e.,
\begin{equation}
    \label{eq:fiscal_externality}
    B_{t}^{fiscal} \coloneqq \tau \cdot \overline{\Delta LI_{t}},
\end{equation}
where $\tau \coloneqq \tau_{pay} + \tau_{inc}$, $\tau_{pay} = 20\%$ is the payroll tax rate that employers must pay to the social security system according to Brazilian Law, and $\tau_{inc} = 8.9\%$ is the average rate of income and social security taxes paid by individuals according to the POF 2017-2018 dataset.\footnote{In the labor income information of POF 2017-2018 dataset, we observe the labor income of each individual, the amount paid in labor income taxes, and the amount of taxes paid to the social security system. Summing all the taxes and all the labor income entries and dividing one by another yields the desired share.}

\medskip 
\noindent\textbf{Total Benefits}. The total benefits of free public transport are the sum of environmental benefits (Equation \eqref{eq:carbon_benefit}) and fiscal benefits (Equation \eqref{eq:fiscal_externality}). Formally, they are defined as
\begin{equation}
    \label{EqTotalBenefits}
    B_{t}^{total} \coloneqq B_{t}^{env} + B_{t}^{fiscal}.
\end{equation}

\begin{figure}[htbp]
\begin{center}
\caption{Annual Benefits of Free Public Transit in Millions of Brazilian reais of 2021}
\label{fig:direct_benefits_policy}
\begin{subfigure}{0.45\linewidth}
\caption{Average Environmental Benefits}
\includegraphics[width=1\linewidth]{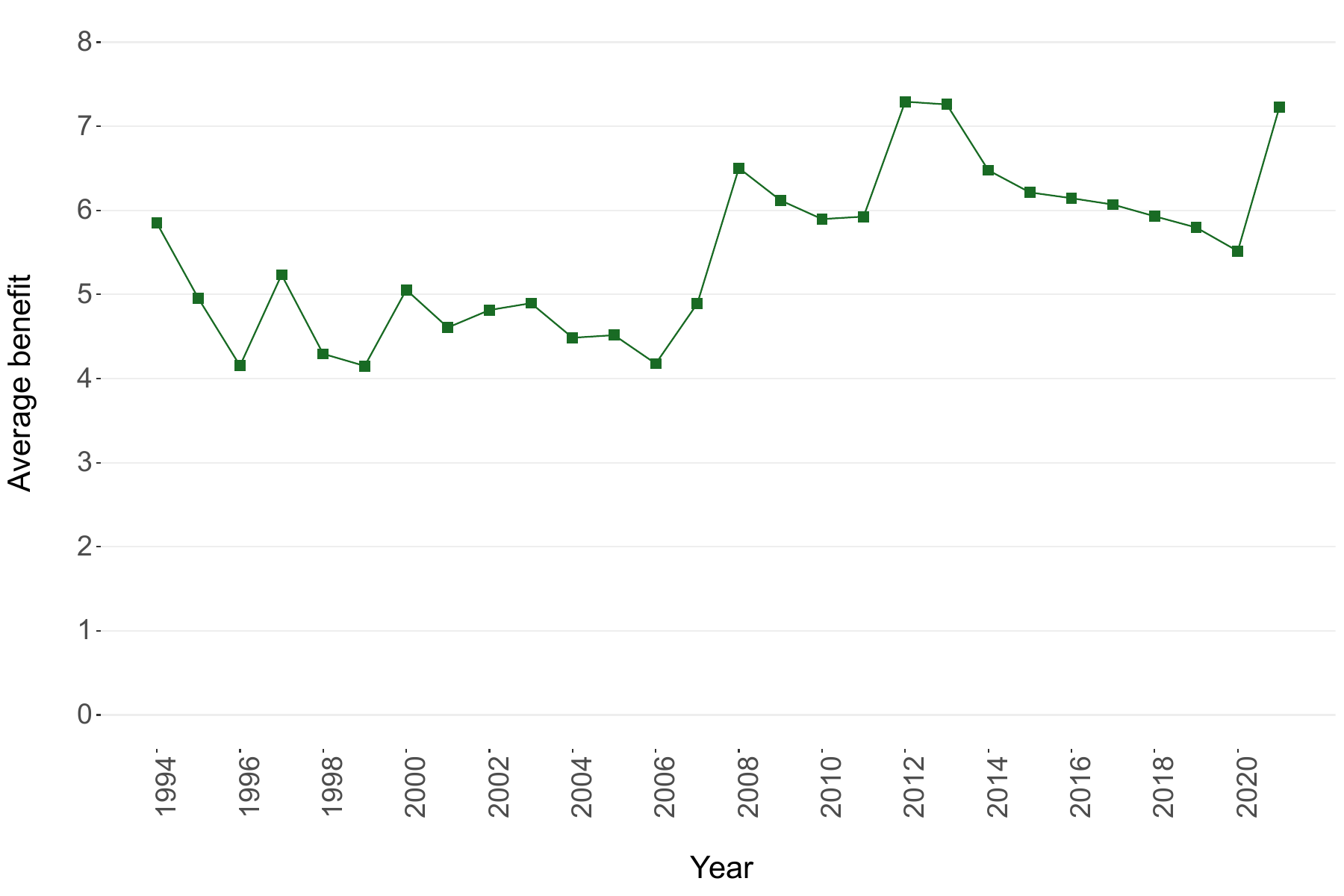}
\label{fig:reducao_media_emissoes}
\end{subfigure} \hfill
\begin{subfigure}{0.45\linewidth}
\caption{Average Fiscal Benefits}
\includegraphics[width=1\linewidth]{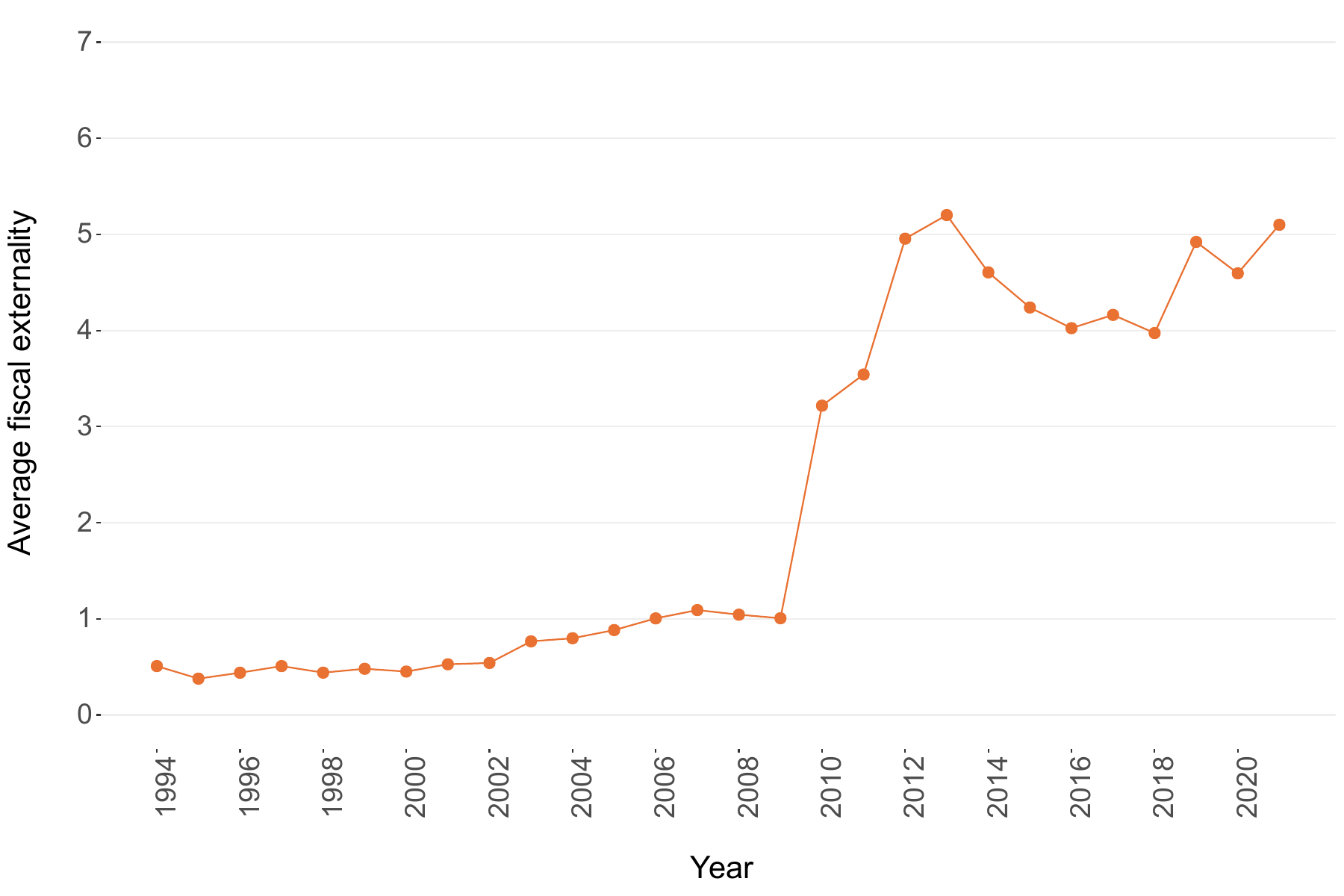}
\label{fig:externalidade_fiscal_media}
\end{subfigure}
\end{center}
\footnotesize{\textit{Notes:} Figure \ref{fig:direct_benefits_policy} presents our estimates of the two types of average benefits of free public transport for the treated municipalities in each year after adoption. Both types of benefits are measured in millions of Brazilian reais of 2021.  Figure \ref{fig:reducao_media_emissoes} shows the estimated monetary value of the environmental benefits due to reduced greenhouse gas emissions (Equation \eqref{eq:carbon_benefit}). Figure \ref{fig:externalidade_fiscal_media} shows the estimated fiscal benefits due to a larger tax revenue from an increased number of formal jobs (Equation \eqref{eq:fiscal_externality}).}
\end{figure}

\begin{figure}[htbp]
\begin{center}
\caption{Annual Benefits of Free Public Transit in Millions of Brazilian Reais of 2021}
\label{fig:average_cost_benefit}
\begin{subfigure}{0.47\linewidth}
\caption{Both Externalities}
\includegraphics[width=1\linewidth]{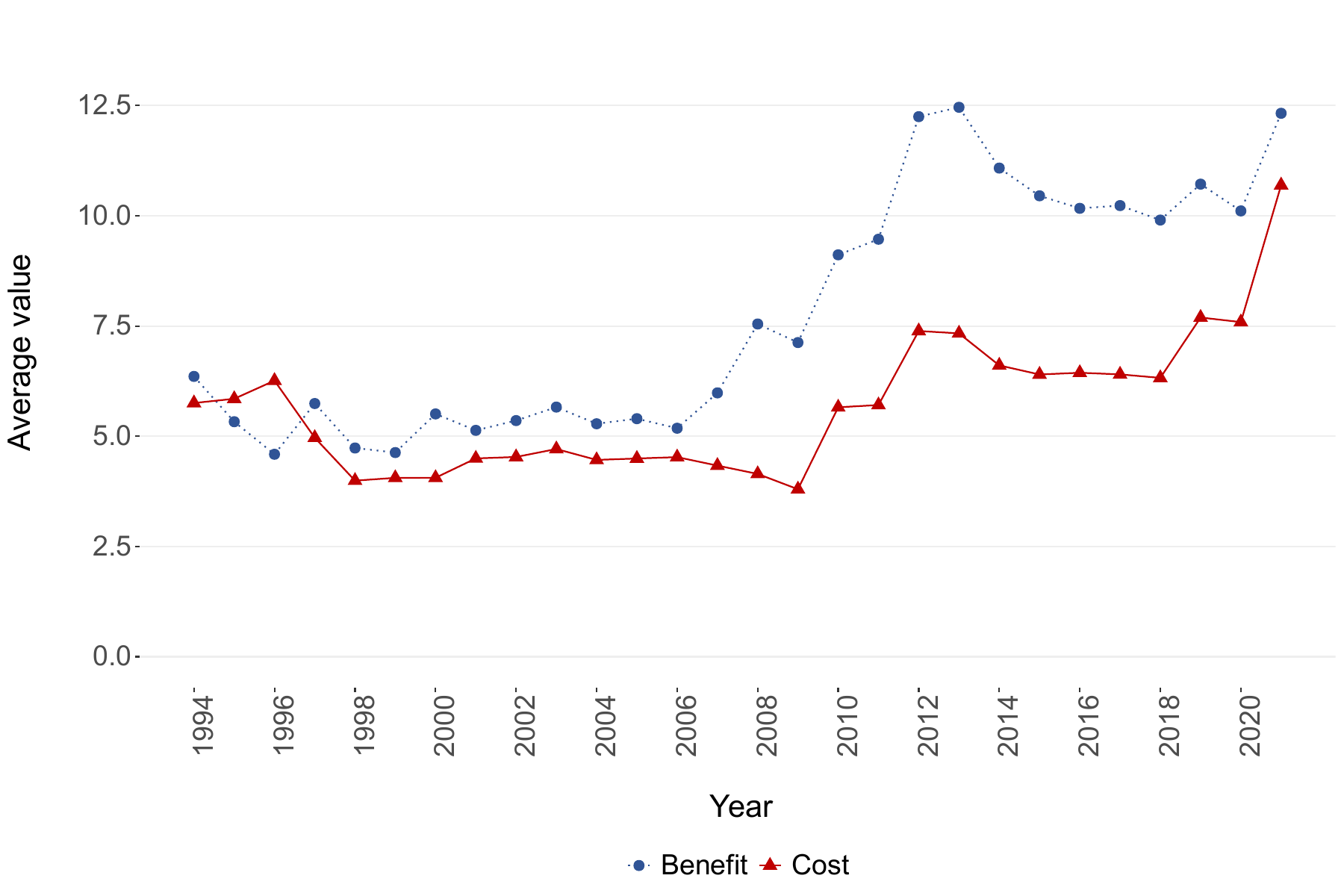}
\label{fig:total_cost_benefit}
\end{subfigure} \hfill
\begin{subfigure}{0.47\linewidth}
\caption{Only Fiscal Externality}
\includegraphics[width=1\linewidth]{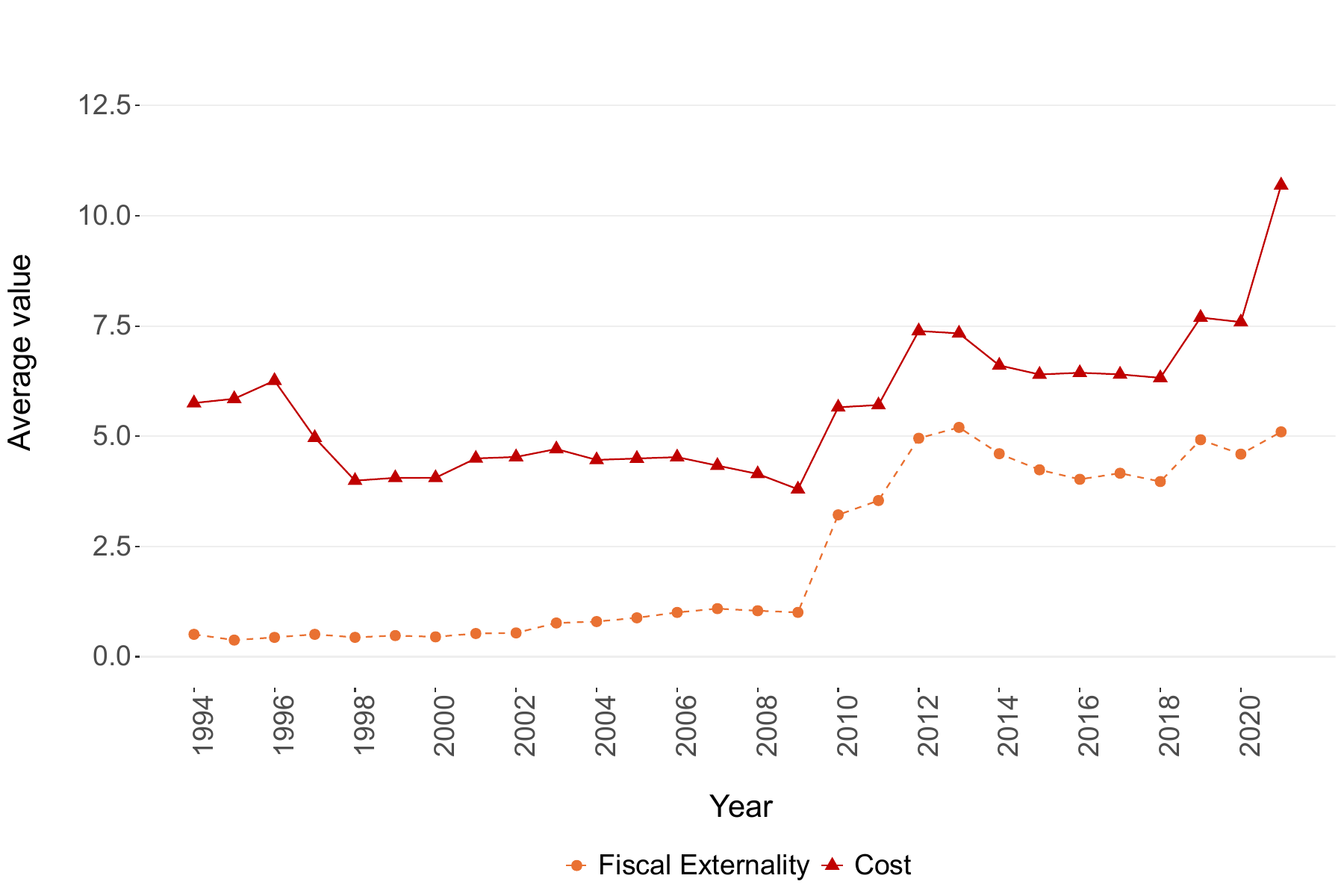}
\label{fig:cost_benefit_only_fiscal}
\end{subfigure}
\end{center}
\footnotesize{\textit{Notes:} This figure presents the average costs and benefits of free public transport for the treated municipalities in each year after adoption. The blue circles show the estimated average total benefit (Equation \eqref{EqTotalBenefits}). Figure (\ref{fig:total_cost_benefit}) compares the policy's cost with the full benefits, comprising both the fiscal and environmental externalities. The red triangles show the estimated average cost of our target policy, i.e., the estimated government expenditures due to larger subsidies to the public transit system. Figure (\ref{fig:cost_benefit_only_fiscal}) compares the policy's cost with only the fiscal externality. All variables are measured in millions of Brazilian Reais of 2021.}
\end{figure}

\medskip 
\noindent\textbf{Results of the Cost-Benefit Analysis}. Figures \ref{fig:direct_benefits_policy} and \ref{fig:average_cost_benefit} present all the results of our cost-benefit analysis of free public transport. Figure \ref{fig:direct_benefits_policy} shows our estimates of the two types of average benefits for the treated municipalities in each year after adoption. Figure \ref{fig:reducao_media_emissoes} shows the estimated monetary value of the environmental benefits due to reduced greenhouse gas emissions (Equation \eqref{eq:carbon_benefit}) while Figure \ref{fig:externalidade_fiscal_media} shows the estimated fiscal benefits due to a larger tax revenue from an increased number of formal jobs (Equation \eqref{eq:fiscal_externality}). Furthermore, the blue dotted circles in Figure \ref{fig:average_cost_benefit} show the estimated average total benefit (Equation \eqref{EqTotalBenefits}) while the red solid triangles in Figure \ref{fig:average_cost_benefit} show the estimated government expenditures due to larger subsidies to the public transit system.

% \begin{figure}[htbp]
% \begin{center}
% \caption{Annual Benefits of Free Public Transit in Millions of Brazilian reais of 2021}
% \label{fig:direct_benefits_policy}
% \begin{subfigure}{0.45\linewidth}
% \caption{Average Environmental Benefits}
% \includegraphics[width=1\linewidth]{Files/reducao_media_emissoes.png}
% \label{fig:reducao_media_emissoes}
% \end{subfigure} \hfill
% \begin{subfigure}{0.45\linewidth}
% \caption{Average Fiscal Benefits}
% \includegraphics[width=1\linewidth]{Files/externalidade_fiscal_media.png}
% \label{fig:externalidade_fiscal_media}
% \end{subfigure}
% \end{center}
% \footnotesize{\textit{Notes:} Figure \ref{fig:direct_benefits_policy} presents our estimates of the two types of average benefits of free public transport for the treated municipalities in each year after adoption. Both types of benefits are measured in millions of Brazilian reais of 2021.  Figure \ref{fig:reducao_media_emissoes} shows the estimated monetary value of the environmental benefits due to reduced greenhouse gas emissions (Equation \eqref{eq:carbon_benefit}). Figure \ref{fig:externalidade_fiscal_media} shows the estimated fiscal benefits due to a larger tax revenue from an increased number of formal jobs (Equation \eqref{eq:fiscal_externality}).}
% \end{figure}

Note that, from 1997 onward, the average total benefits of free public transit are greater than its average costs. However, neither type of benefit is sufficient to surpass its costs separately. This insufficiency result matters because the estimated environmental benefits are based on a global externality, and local governments are not automatically compensated for them due to the lack of carbon markets pricing them. In addition, in countries operating in fiscal federalism, some taxes are accrued by the national government, which further reduces the incentives of local governments. Importantly, all taxes included in our fiscal benefit estimates are accrued by the federal government, while local governments pay all public transport expenditures.\footnote{Local governments in Brazil collect three types of taxes, which corresponds to 5,5\% of all public sector revenue. Taxes collected by the national government correspond to 70\% of the public sector revenue.} Consequently, cost-effective policies that reduce greenhouse gas emissions and increase tax revenue may not be adopted by municipalities because they might not have adequate incentives.

% \begin{figure}[htb]
% \begin{center}
% \caption{Average Costs and Benefits of Free Public Transit by Year in Millions of Brazilian reais of 2021}
% \label{fig:average_cost_benefit}
% \includegraphics[width=0.8\linewidth]{Files/custo_beneficio_medio.png}
% \end{center}
% \footnotesize{\textit{Notes:} This figure presents the average costs and benefits of free public transport for the treated municipalities in each year after adoption. The blue circles show the estimated average total benefit (Equation \eqref{EqTotalBenefits}). The red triangles show the estimated average cost of our target policy, i.e., the estimated government expenditures due to larger subsidies to the public transit system. Both variables are measured in millions of Brazilian reais of 2021.}
% \end{figure}

One caveat of our analysis is that the estimated total benefits may need to include other components. First, we are likely underestimating the fiscal benefits of the policy. For instance, we only considered income and social security taxes when estimating the fiscal benefits. We did so because increasing their revenue is a direct consequence of increasing formal jobs. However, there are other fiscal benefits that we did not include in our analysis. For example, if more formal jobs are associated with increased consumption, the revenue from sales taxes will also rise, elevating the fiscal benefits even more. Second, the policy may also generate health benefits by reducing air pollution ($PM_{2.5}$ particles generated by combustion emissions).

Another caveat of our analysis is the likely underestimation of costs because the policy tends to generate over-consumption. To understand the severity of issue, we start by taking the time average of the benefits and costs of free public transport. We find that, if the free-fare public transit increased transport demand by 43\%, the estimated average total benefit would still be greater than the estimated average cost. To understand if free-fare policies are likely to increase transport demand by more than 43\%, we look at the literature studying free-pass interventions. For instance, \cite{Bull2021}, investigating a lower dosage intervention lasting two weeks in Chile, indicate an increase in overall travel of 12\%. \cite{Brough2022}, studying a higher-dosage intervention lasting six months in the U.S., point out increases in the number of trips by 250--300\%. If the increase in trips is closer to the higher-dosage intervention, the costs of our policy would be greater than its aggregate benefits.

%%%%%%%%%%%%%%%%%%%%%%%%%%%%%%%%%%%%%%%%
\section{Concluding Remarks}\label{sec:conclusion}
%%%%%%%%%%%%%%%%%%%%%%%%%%%%%%%%%%%%%%%%

This paper studies the effects of a free transport policy implemented by Brazilian municipalities. We assess how this policy that increases connectivity by providing free bus fares for all residents affects labor market and environmental outcomes. In the empirical analysis, we follow a staggered difference-in-differences approach, using the program rollout to contrast the outcomes of municipalities implementing the policy against comparable never-treated cities. 

Our results show that the policy positively affects employment and decreases GHG emissions. Understanding which transport policy is able to create economic and environmental co-benefits is a policy-relevant issue in the context of climate change and the ambitious emission reduction targets of the Paris Agreement. Our findings suggest that transport policies---beyond electrification and biofuels---can reduce GHG emissions. Therefore, a combination of transport policies may be used in climate mitigation strategies and transitioning to a low-carbon economy with sustainable net-zero emissions. Our findings provide new evidence to the literature and are consistent with the IPCC indicating free-fare public transport as a potentially relevant institutional-led intervention \citep{jaramillo2022transport}.

Since local governments can implement several alternative transportation policies, assessing the costs and benefits of policies that provide high subsidies is relevant. Another implication is that climate mitigation policies may be costly, and cost-benefit analyses must consider both local and global benefits. When a policy is only cost-effective when one considers the benefit from reducing a type of global externality (and there is no price for compensating for this reduction), the take-up of such transport policies may be lower than in counterfactual scenarios with carbon markets pricing externalities. Finally, it is essential to consider the different channels through which the transport policy affects the economy, including those related to the composition of economic activity and structural transformation in response to policies.

\bigskip

\singlespacing
\bibliographystyle{aer}
\bibliography{Bib}

\appendix

\pagenumbering{arabic}% resets `page` counter to 1
\renewcommand*{\thepage}{A-\arabic{page}}
\setcounter{page}{1}
\pagestyle{online}

\setcounter{table}{0}
\renewcommand{\thetable}{A-\arabic{table}}

\renewcommand{\theequation}{\Alph{section}.\arabic{equation}}

\setcounter{figure}{0}
\renewcommand{\thefigure}{A-\arabic{figure}}

\newcommand{\diag}{\mathop{\mathrm{diag}}}

\newgeometry{top=1in,left=1in,right=1in}

\emptythanks
\title{\Large Online Appendix to ``\Title''}

\author{\large Mateus Rodrigues, Daniel Da Mata and Vitor Possebom}
\date{\normalsize \today}

\maketitle

%content

{
  \hypersetup{linkcolor=black}
\startcontents[section]
\printcontents[section]{l}{1}{\setcounter{tocdepth}{3}}
}

\onehalfspacing

\numberwithin{table}{section}
\numberwithin{figure}{section}

\setlength\abovedisplayskip{5pt}
\setlength\belowdisplayskip{5pt}

\thispagestyle{online}\clearpage

\subsectionfont{\normalsize}

%%%%%%%%%%%%%%%%%%%%%%%%%%%%%%%%%%%%%%%%
\section{Heterogeneity Analysis: Average Effects for each Adoption-year Group of Municipalities}\label{sec:results-group}
%%%%%%%%%%%%%%%%%%%%%%%%%%%%%%%%%%%%%%%%

This section shows the average effects for municipalities adopting free public transport for the first time in each possible year across all their post-treatment years (Equation \eqref{eq:treatment_effect_group}). These target parameters are useful to understand if early adopters are positively selected in comparison to late adopters.

\begin{figure}[htbp]
\begin{center}
\caption{Average Effects for each Adoption-year Group}
\label{fig:gwp_100_ar5_ghg_vinculos_ativos_group_pandemia}
\begin{subfigure}{0.8\linewidth}
\caption{Outcome Variable: Employment}
\includegraphics[width=1\linewidth]{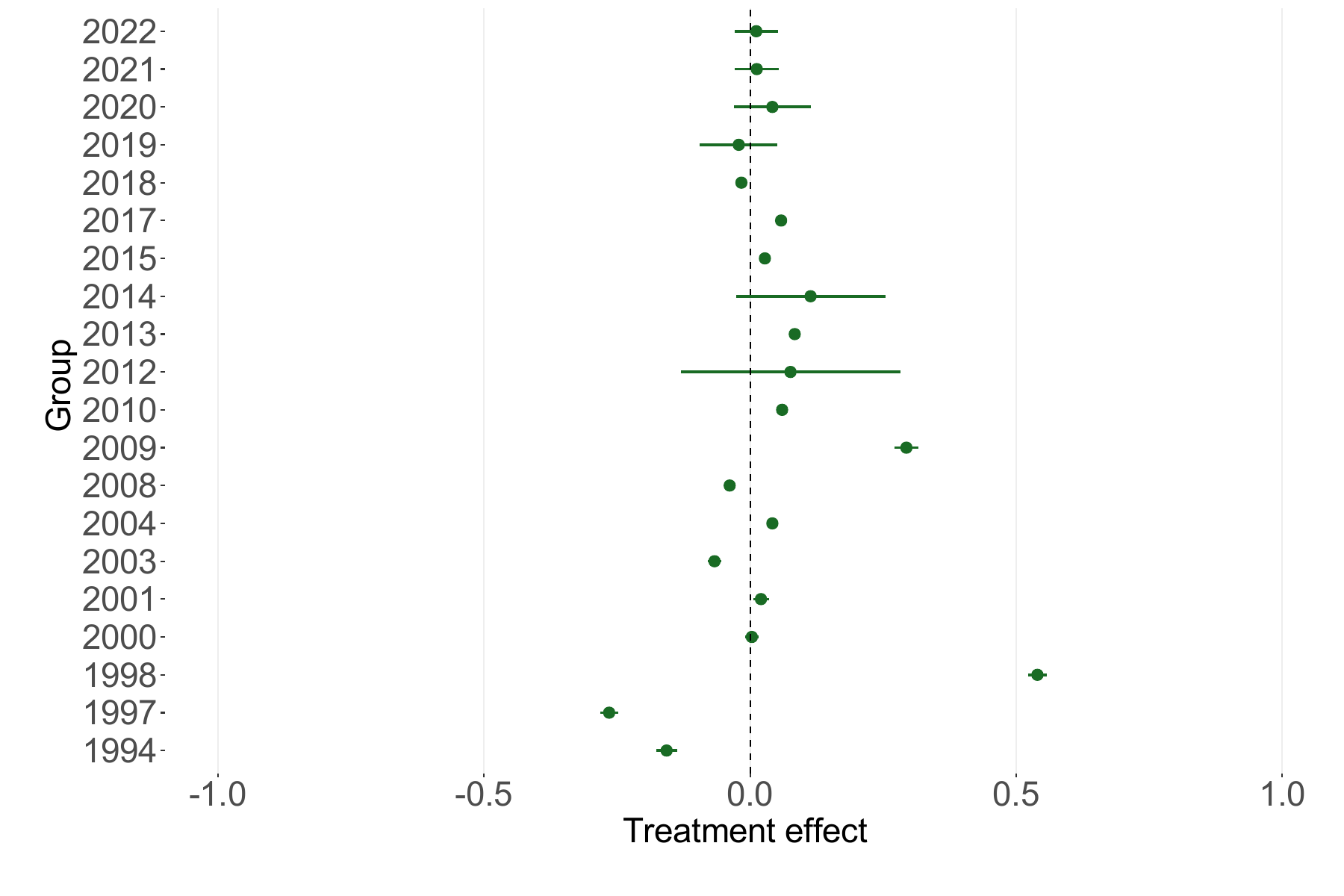}
\label{fig:vinculos_ativos_group_pandemia}
\end{subfigure} \\
\begin{subfigure}{0.8\linewidth}
\caption{Outcome Variable: Greenhouse Gas Emissions}
\includegraphics[width=1\linewidth]{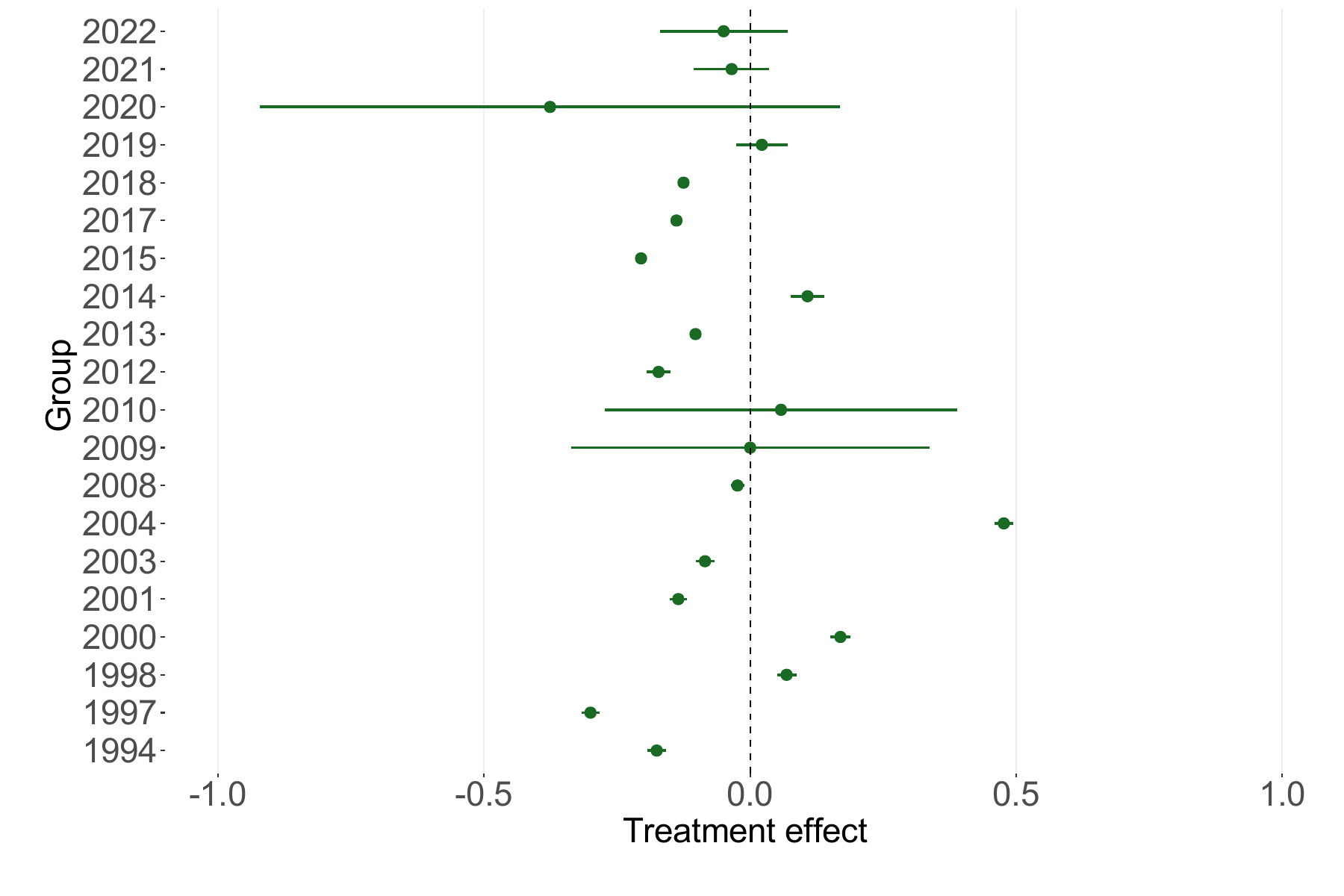}
\label{fig:gwp_100_ar5_ghg_group_pandemia}
\end{subfigure}
\end{center}
\footnotesize{\textit{Notes:} Figure \ref{fig:gwp_100_ar5_ghg_vinculos_ativos_group_pandemia} presents estimates of the average effects for municipalities adopting free public transport for the first time in each year $g$ across all their post-treatment years (Equation \eqref{eq:treatment_effect_group}). The outcome variables are the natural logarithm of the formally employed individuals in each municipality (Figure \ref{fig:vinculos_ativos_group_pandemia}) and the natural logarithm of CO\textsubscript{2}-equivalent emissions in each municipality (Figure \ref{fig:gwp_100_ar5_ghg_group_pandemia}). Vertical lines represent point-wise 90\%-confidence intervals based on standard errors clusterized at the municipality level. These results are based on the doubly-robust estimator proposed by \cite{callaway2021difference}.}
\end{figure}

Appendix Figure \ref{fig:gwp_100_ar5_ghg_vinculos_ativos_group_pandemia} presents estimates of the average effects for municipalities taking the treatment for the first time in each year $g$ across all their post-treatment years (Equation \eqref{eq:treatment_effect_group}). The outcome variables are the natural logarithm of the formally employed individuals (Appendix Figure \ref{fig:vinculos_ativos_group_pandemia}) and the natural logarithm of CO\textsubscript{2}-equivalent emissions (Appendix Figure \ref{fig:gwp_100_ar5_ghg_group_pandemia}). Vertical lines represent point-wise 90\%-confidence intervals based on standard errors clustered at the municipality level. These results are based on a doubly-robust estimator.

Note that, when we use employment as the outcome variable (Appendix Figure \ref{fig:vinculos_ativos_group_pandemia}), 14 out of 20 estimates of the average treatment effect for adoption-year groups are positive, and eight out of them are statistically significant. Moreover, there is weakly suggestive evidence that early adopters of free public transport might be negatively selected in comparison with late adopters.

Furthermore, observe that, when we use greenhouse gas emissions as the outcome variable (Appendix Figure \ref{fig:gwp_100_ar5_ghg_group_pandemia}), 13 out of 20 estimates of the average treatment effect for adoption-year groups are negative, and 10 out of them are statistically significant. In addition, there is weakly suggestive evidence that early adopters have smaller decreases (or even increases) in emissions in comparison with late adopters.

\newpage

%%%%%%%%%%%%%%%%%%%%%%%%%%%%%%%%%%%%%%%%
\section{Robustness Check: Excluding Years During the Covid-19 Pandemic}\label{appendix_covid}
%%%%%%%%%%%%%%%%%%%%%%%%%%%%%%%%%%%%%%%%

Although we included the years during the COVID-19 Pandemic in our main set of results (Section \ref{sec:results}), one may wonder if the unique dynamics of this period may have impacted our estimates. This concern may be specially important considering that there is a large number of municipalities that adopted free public transport policies during this period. In particular, employment location choices may have been strongly impacted by the increase in remote work.

To overcome any concerns with respect to the uniqueness of the COVID-19 Pandemic years, we re-estimate all our target parameters removing the years between 2020 and 2022 from our sample. This appendix presents the results using this smaller sample. Importantly, these results are very similar to the ones in the main text, showing the robustness of our results to different sampling periods.

Table \ref{tab:overall_att_estimates_seeg_ghg_employment} presents the estimates of the Average Effect of Adopting Free Public Transit Across All Treated Municipalities during the Pre-Pandemic Sample. It focuses on two main outcomes: the natural logarithm of the formally employed individuals in each municipality (Column (1)) and the natural logarithm of net CO\textsubscript{2}-equivalent emissions in each municipality (Column (2)). These results are based on the doubly-robust estimator proposed by \cite{callaway2021difference} and standard errors are clustered at the municipality level.

\begin{table}[htbp]
\begin{center}
\caption{\label{tab:overall_att_estimates_seeg_ghg_employment}Average Effect of Adopting Free Public Transit Across All Treated Municipalities}
\begin{tabular}{lcc}
\toprule \toprule
& \multicolumn{2}{c}{Outcome Variable} \\ \hline
 & Employment & GHG emissions \\
 & (1) & (2) \\
\midrule
ATT (Summary) & $0.038^{**}$ & $0.004^{}$ \\
 & (0.016) & (0.028)\\
\midrule
Treated units & 26 & 26\\
Groups & 17 & 17\\
Units & 2242 & 2204\\
\bottomrule \bottomrule
\end{tabular}
\end{center}
\footnotesize{\textit{Notes:} This table presents the estimates of the average effect of adopting free public transit across all municipalities that ever took treatment (Equation \eqref{eq:overall_treatment_effect}). The outcome variables are the natural logarithm of the formally employed individuals in each municipality (Column (1)) and the natural logarithm of net CO\textsubscript{2}-equivalent emissions in each municipality (Column (2)). These results are based on the doubly-robust estimator proposed by \cite{callaway2021difference}. Standard errors are reported in parentheses and are clustered at the municipality level. At the bottom, we also report the total number of municipalities in our samples, the number of treated municipalities and the number of adoption-year groups. Significance levels are denoted as follows: $^{***}p<0.01$; $^{**}p<0.05$; $^{*}p<0.1$.}
\end{table}

Similarly to our main results (Table \ref{tab:overall_att_estimates_seeg_ghg_employment_pandemic}), these general-purpose summary measures are consistent with absolute decoupling. In particular, we find that emissions are unaffected by free public transport policies while employment is positively impacted by this policy.

\begin{figure}[htbp]
\begin{center}
\caption{Average Effects for each Length of Exposure to the Treatment --- Pre-Pandemic Sample}
\label{fig:gwp_100_ar5_ghg_vinculos_ativos_dynamic}
\begin{subfigure}{0.75\linewidth}
\caption{Outcome Variable: Employment}
\includegraphics[width=1\linewidth]{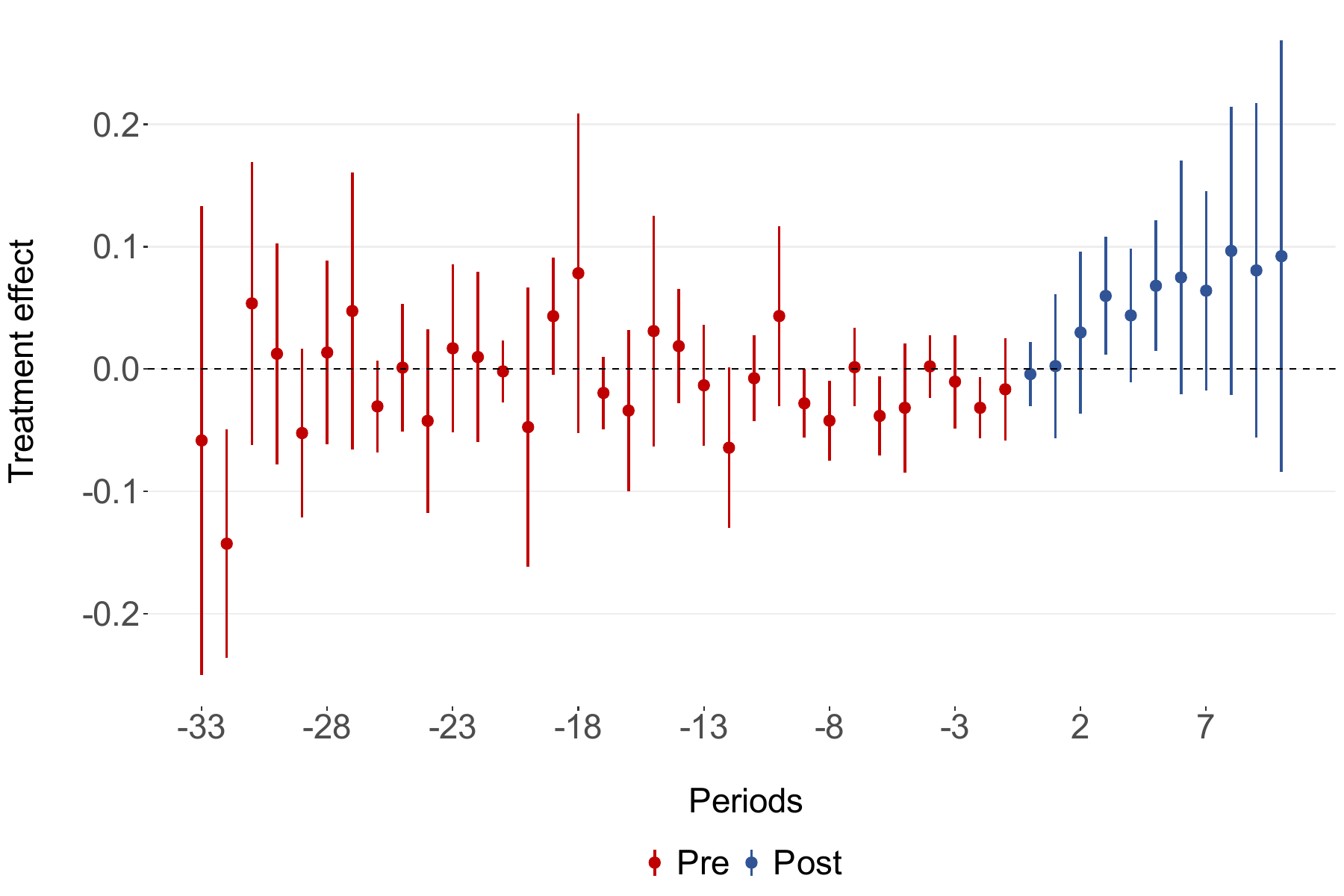}
\label{fig:vinculos_ativos_dynamic}
\end{subfigure} \\
\begin{subfigure}{0.75\linewidth}
\caption{Outcome Variable: Greenhouse Gas Emissions}
\includegraphics[width=1\linewidth]{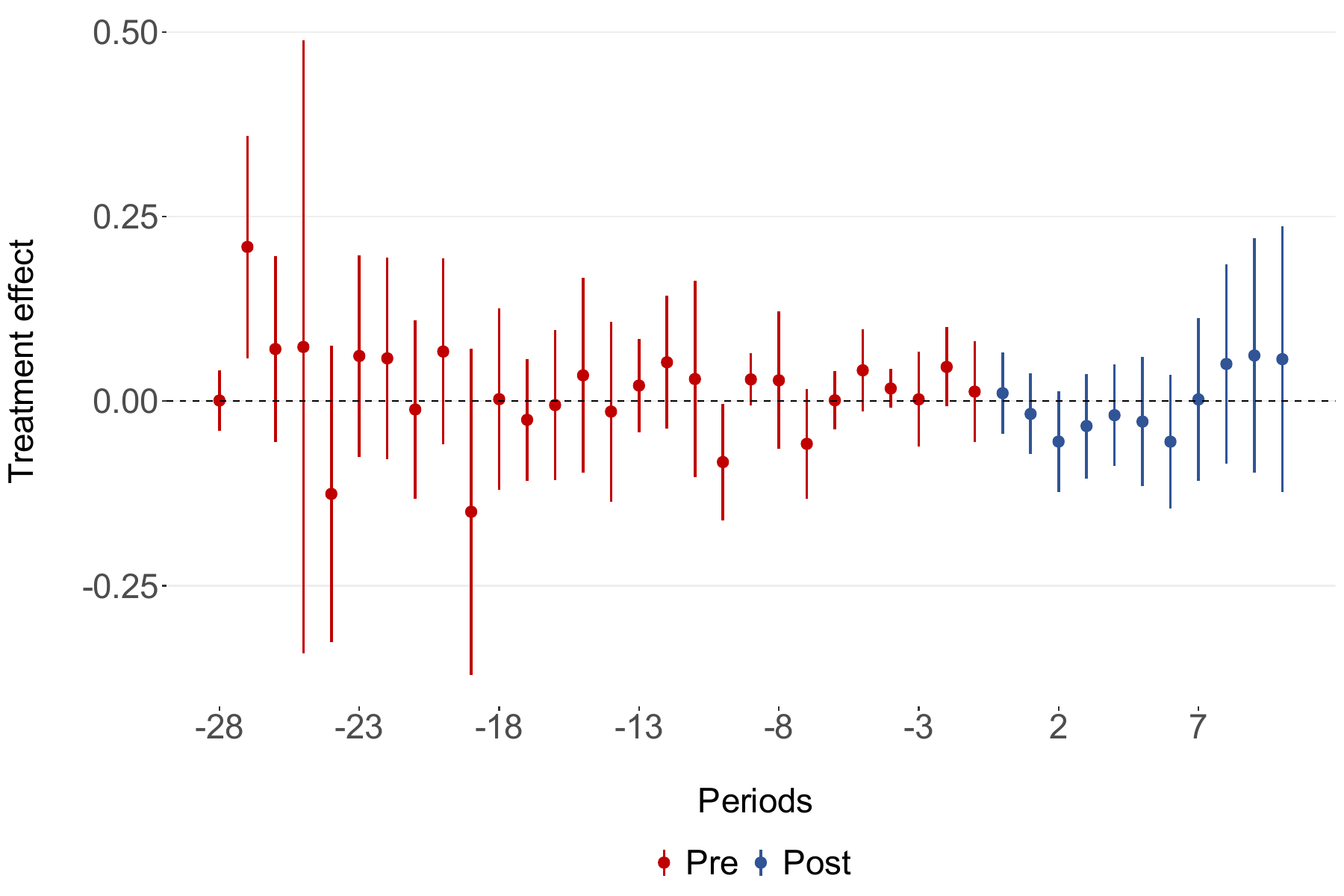}
\label{fig:gwp_100_ar5_ghg_dynamic}
\end{subfigure}
\end{center}
\footnotesize{\textit{Notes:} Figure \ref{fig:gwp_100_ar5_ghg_vinculos_ativos_dynamic} presents estimates of the average effect of adopting free public transit $e$ years after adoption across all municipalities that are ever observed to have taken the treatment for exactly $e$ periods (Equation \eqref{eq:treatment_effect_length_exposure}) before the COVID-19 Pandemic. The outcome variables are the natural logarithm of the formally employed individuals in each municipality (Figure \ref{fig:vinculos_ativos_dynamic}) and the natural logarithm of net CO\textsubscript{2}-equivalent emissions in each municipality (Figure \ref{fig:gwp_100_ar5_ghg_dynamic}). Vertical lines represent point-wise 90\%-confidence intervals based on standard errors clusterized at the municipality level. These results are based on the doubly-robust estimator proposed by \cite{callaway2021difference} using a varying base period for the pre-treatment placebo estimates (in red). Post-treatment estimates are reported in blue.}
\end{figure}

Moreover, Figure \ref{fig:gwp_100_ar5_ghg_vinculos_ativos_dynamic} presents estimates of the average effect of adopting free public transport $e$ years after adoption across all municipalities that are ever observed to have taken the treatment for exactly $e$ periods (Equation \eqref{eq:treatment_effect_length_exposure}) before the COVID-19 Pandemic. The outcome variables are the natural logarithm of the formally employed individuals in each municipality (Figure \ref{fig:vinculos_ativos_dynamic_pandemia}) and the natural logarithm of net CO\textsubscript{2}-equivalent emissions in each municipality (Figure \ref{fig:gwp_100_ar5_ghg_dynamic_pandemia}). Vertical lines represent point-wise 90\%-confidence intervals based on standard errors clusterized at the municipality level. While post-treatment effects are reported in blue, pre-treatment placebo estimates (varying base period) are reported in red. These results are based on the doubly-robust estimator proposed by \cite{callaway2021difference}.

These results for the pre-pandemic period are similar to our main results (Figure \ref{fig:gwp_100_ar5_ghg_vinculos_ativos_dynamic_pandemia}). First, there is no evidence of anticipation effects or violations of the parallel trends assumption for either of our outcome variables. Additionally, we find positive effects when we use employment as our outcome variable and null effects when we use greenhouse gas emissions as our outcome variable.

Lastly, Figure \ref{fig:gwp_100_ar5_ghg_vinculos_ativos_group_pandemia} presents estimates of the average effects for municipalities taking the treatment for the first time in each year $g$ across all their post-treatment years (Equation \eqref{eq:treatment_effect_group}) before the COVID-19 Pandemic. The outcome variables are the natural logarithm of the formally employed individuals in each municipality (Figure \ref{fig:vinculos_ativos_group_pandemia}) and the natural logarithm of net CO\textsubscript{2}-equivalent emissions in each municipality (Figure \ref{fig:gwp_100_ar5_ghg_group_pandemia}). Vertical lines represent point-wise 90\%-confidence intervals based on standard errors clusterized at the municipality level. These results are based on the doubly-robust estimator proposed by \cite{callaway2021difference}.

\begin{figure}[htbp]
\begin{center}
\caption{Average Effects for each Adoption-year Group --- Pre-Pandemic Sample}
\label{fig:gwp_100_ar5_ghg_vinculos_ativos_group}
\begin{subfigure}{0.8\linewidth}
\caption{Outcome Variable: Employment}
\includegraphics[width=1\linewidth]{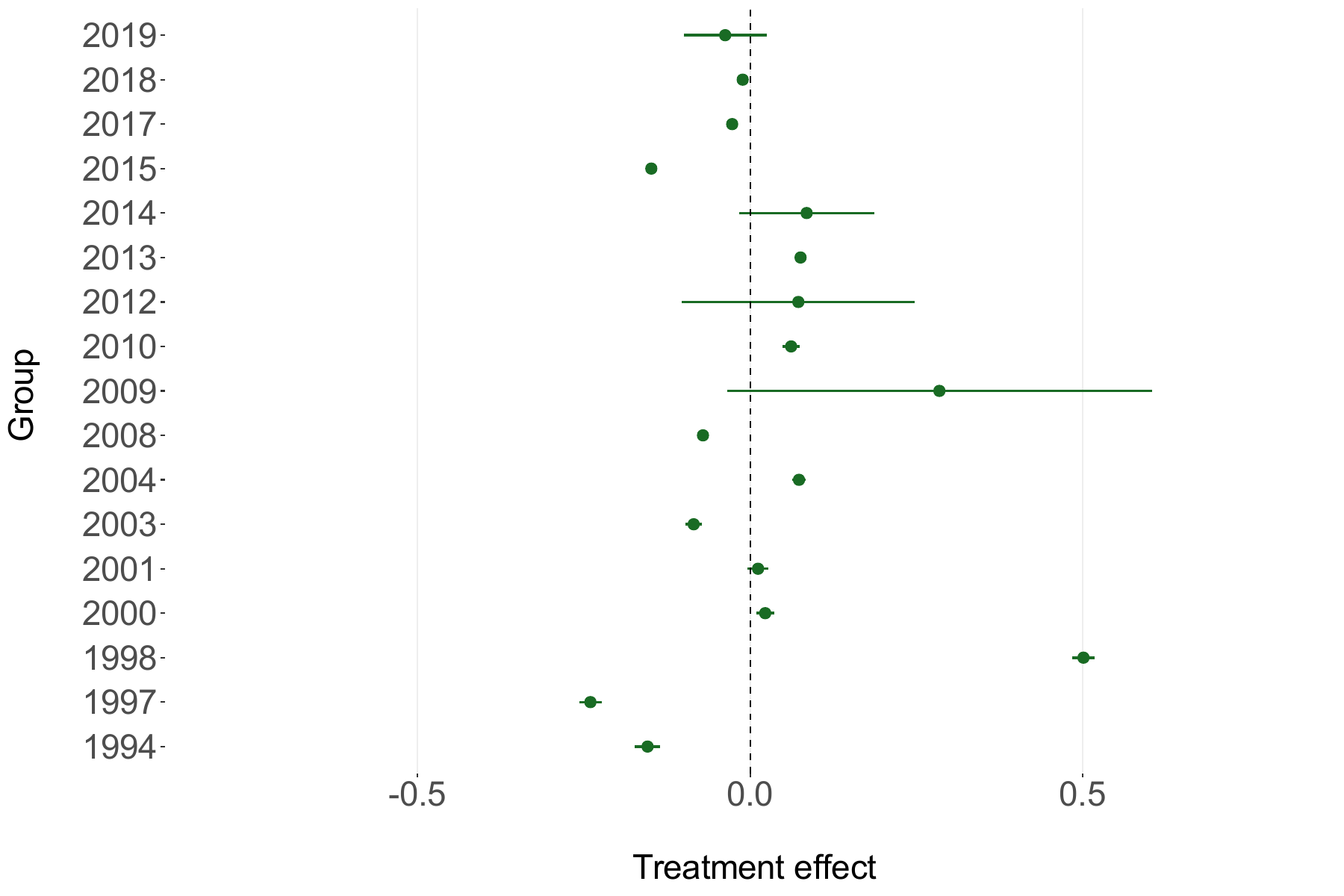}
\label{fig:vinculos_ativos_group}
\end{subfigure} \\
\begin{subfigure}{0.8\linewidth}
\caption{Outcome Variable: Greenhouse Gas Emissions}
\includegraphics[width=1\linewidth]{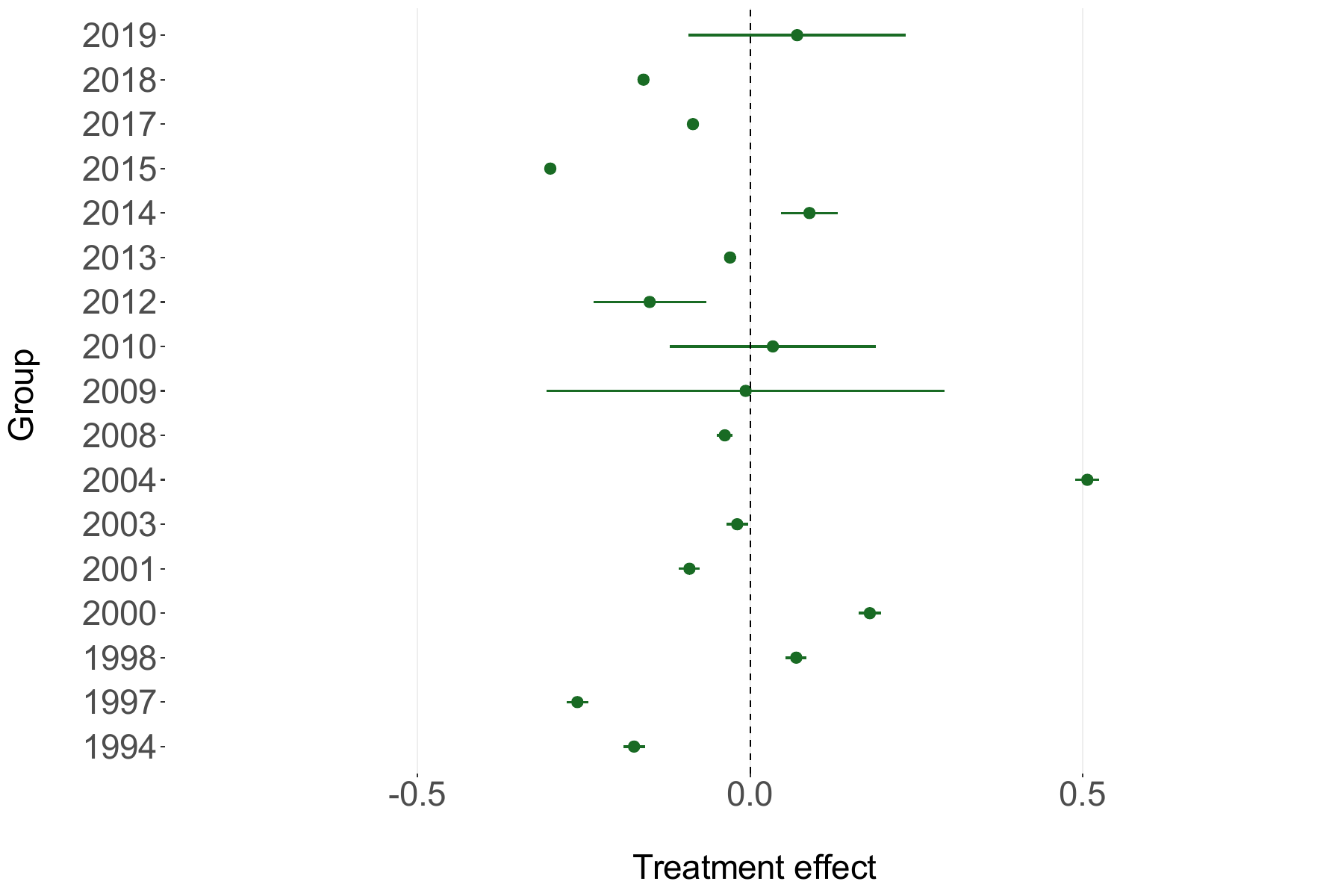}
\label{fig:gwp_100_ar5_ghg_group}
\end{subfigure}
\end{center}
\footnotesize{\textit{Notes:} Figure \ref{fig:gwp_100_ar5_ghg_vinculos_ativos_group} presents estimates of the average effects for municipalities adopting free public transit for the first time in each year $g$ across all their post-treatment years (Equation \eqref{eq:treatment_effect_group}) before the COVID-19 Pandemic. The outcome variables are the natural logarithm of the formally employed individuals in each municipality (Figure \ref{fig:vinculos_ativos_group}) and the natural logarithm of net CO\textsubscript{2}-equivalent emissions in each municipality (Figure \ref{fig:gwp_100_ar5_ghg_group}). Vertical lines represent point-wise 90\%-confidence intervals based on standard errors clusterized at the municipality level. These results are based on the doubly-robust estimator proposed by \cite{callaway2021difference}.}
\end{figure}

Once more, these results for the pre-pandemic period are similar to our main results (Figure \ref{fig:gwp_100_ar5_ghg_vinculos_ativos_group_pandemia}). We emphasize that all adoption-year group estimates have similar magnitude and direction for both sampling periods, illustrating the robustness of our results to different sampling periods.

\newpage

%%%%%%%%%%%%%%%%%%%%%%%%%%%%%%%%%%%%%%%%
\section{Extra Tables and Figures}\label{appendix_tables_figures}
%%%%%%%%%%%%%%%%%%%%%%%%%%%%%%%%%%%%%%%%

\begin{figure}[H]
\begin{center}
\caption{Average Effects for each Length of Exposure to the Treatment}
\label{fig:automovel_vendas_combustiveis_dynamic}
\begin{subfigure}{0.45\linewidth}
\caption{Outcome Variable: Stock of Cars}
\includegraphics[width=1\linewidth]{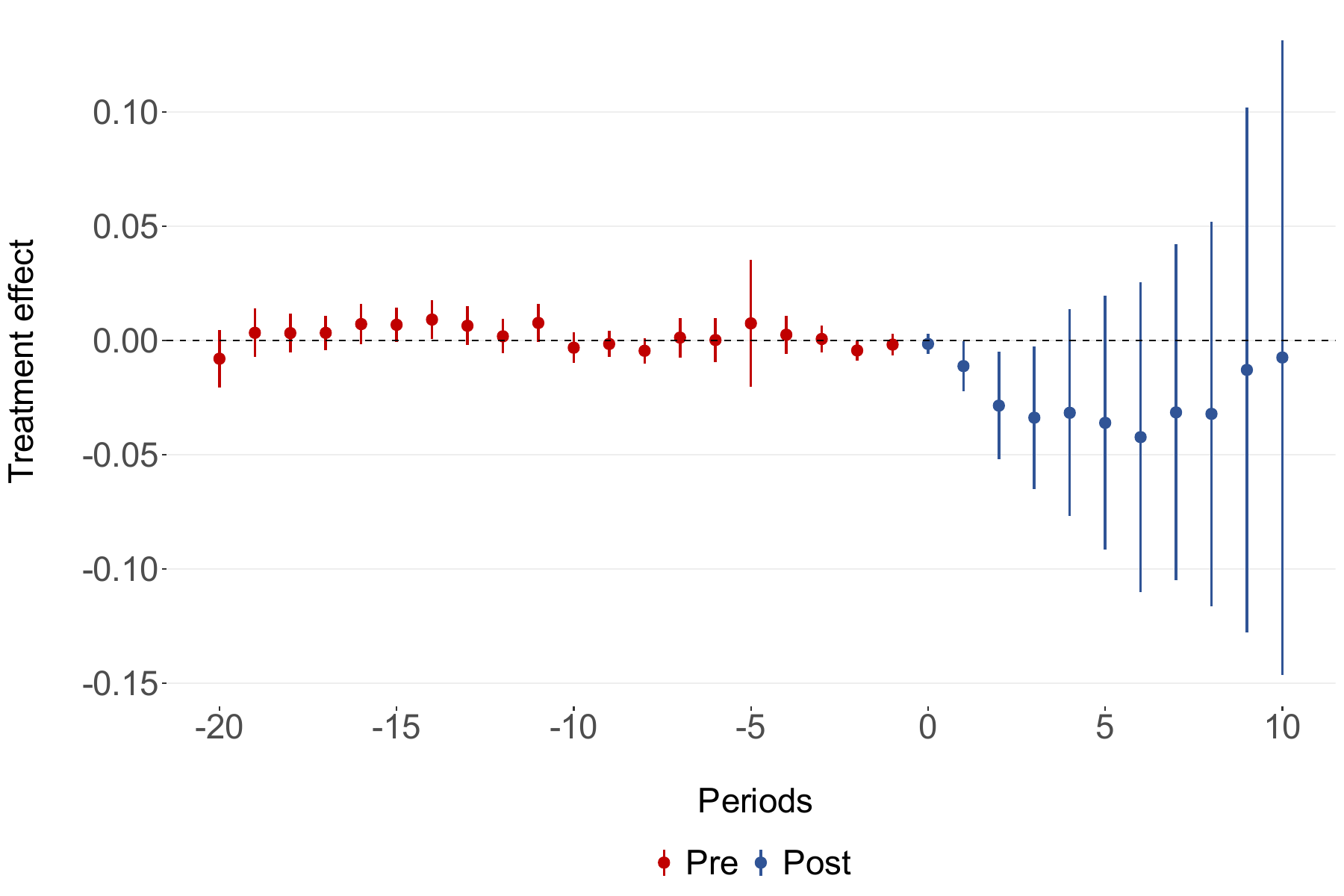}
\label{fig:automovel_dynamic_pandemia}
\end{subfigure}
\begin{subfigure}{0.45\linewidth}
\caption{Outcome Variable: Gasoline Sales}
\includegraphics[width=1\linewidth]{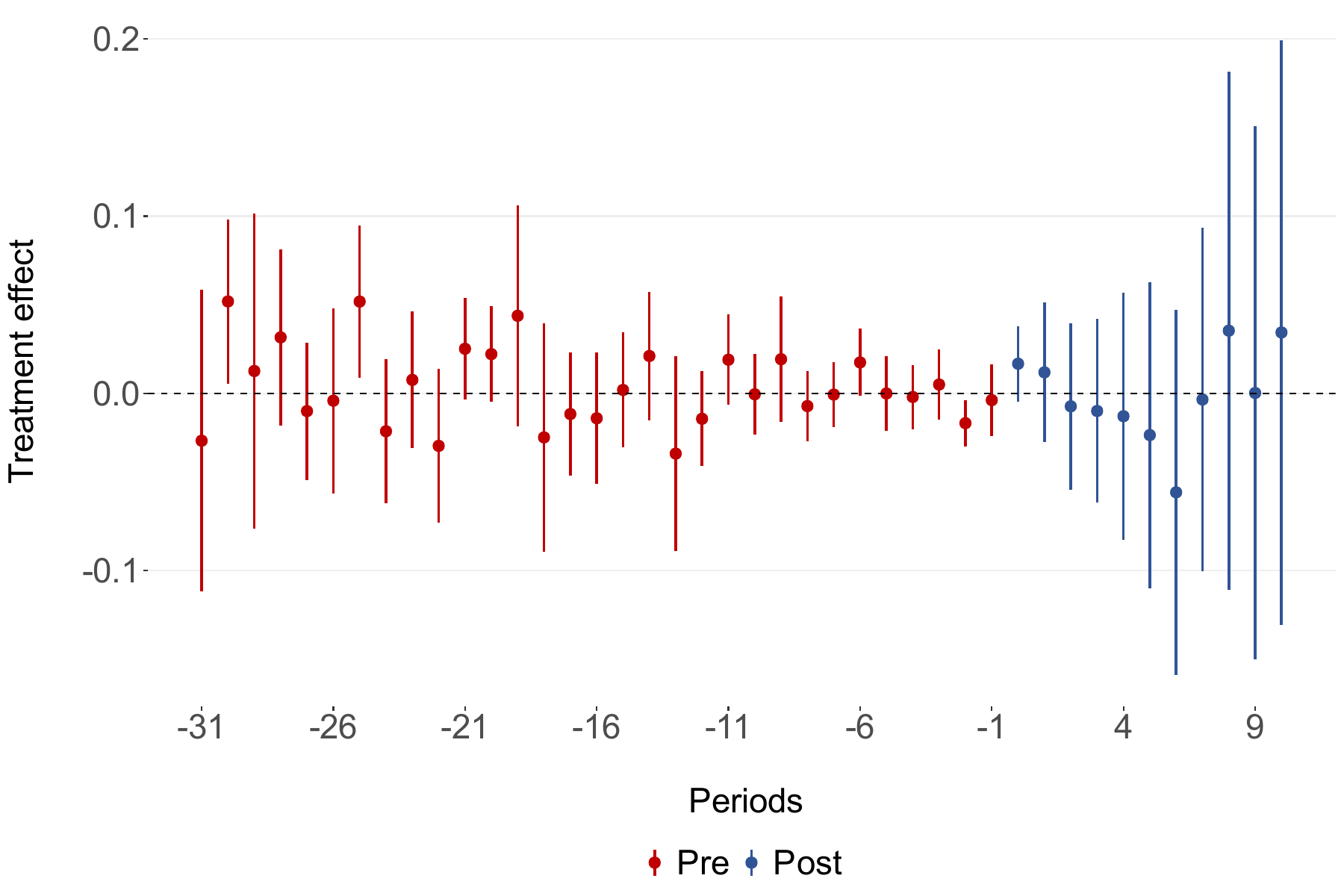}
\label{fig:vendas_gasolina_dynamic_pandemia}\end{subfigure} \\
\begin{subfigure}{0.45\linewidth}
\caption{Outcome Variable: Ethanol Sales}
\includegraphics[width=1\linewidth]{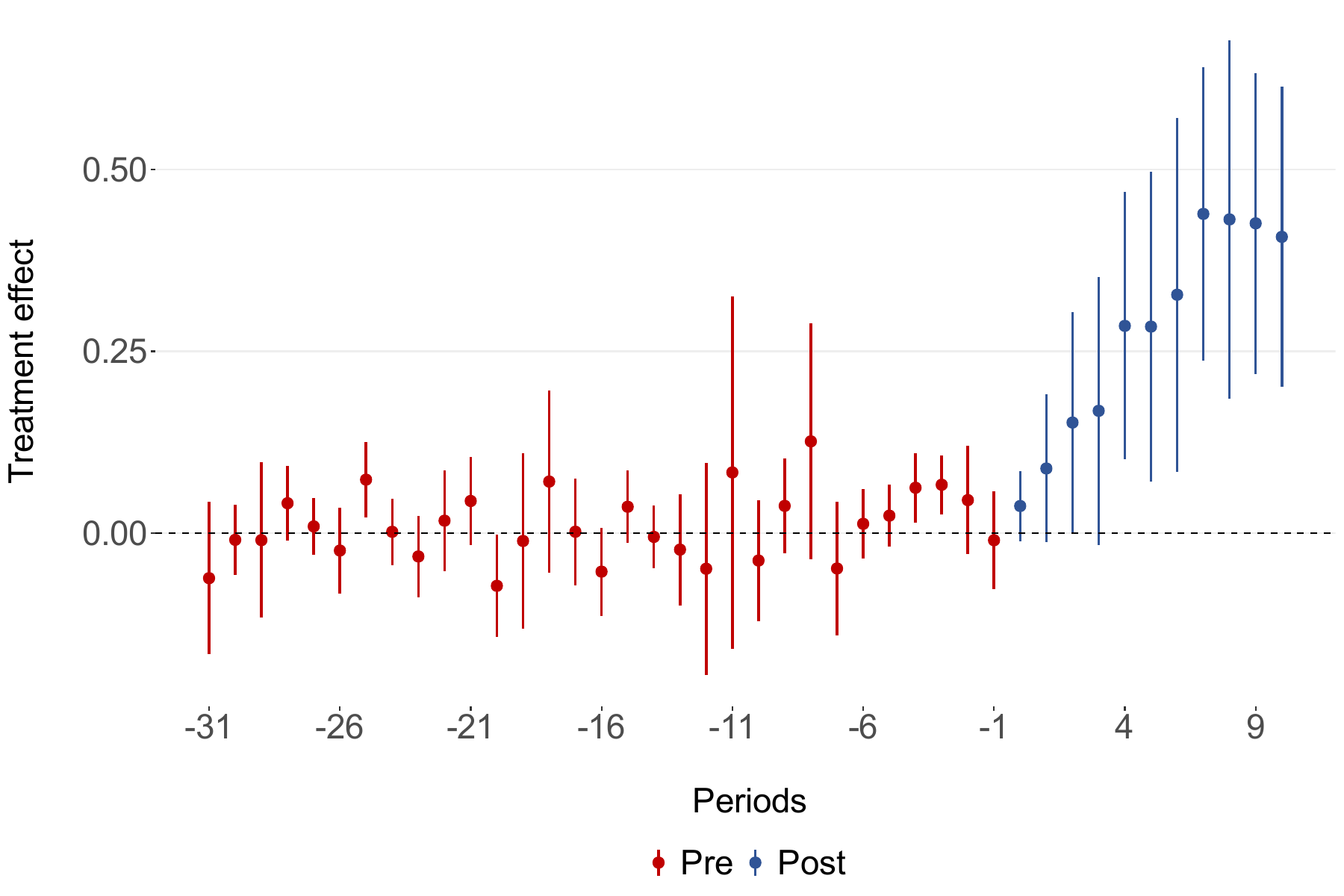}
\label{fig:vendas_etanol_dynamic_pandemia}
\end{subfigure} \\
\end{center}
\footnotesize{\textit{Notes:} Figure \ref{fig:automovel_vendas_combustiveis_dynamic} presents estimates of the average effect of adopting free public transit $e$ years after adoption across all municipalities that are ever observed to have taken the treatment for exactly $e$ periods (Equation \eqref{eq:treatment_effect_length_exposure}). The outcome variables are the natural logarithm of the stock of automobiles in each municipality (Figure \ref{fig:automovel_dynamic_pandemia}), the natural logarithm of sales of gasoline in each municipality (Figure \ref{fig:vendas_gasolina_dynamic_pandemia}), and the natural logarithm of sales of ethanol in each municipality (Figure \ref{fig:vendas_etanol_dynamic_pandemia}). Vertical lines represent uniform 90\%-confidence intervals based on standard errors clusterized at the municipality level. These results are based on the doubly-robust estimator proposed by \cite{callaway2021difference} using a varying base period for the pre-treatment placebo estimates (in red). Post-treatment estimates are reported in blue.}
\end{figure}

\begin{figure}[H]
\begin{center}
\caption{Average Effects for each Length of Exposure to the Treatment \\ Outcome Variable: Formal Employment Level by Sector}
\label{fig:vinculos_ativos_setor_dynamic}
\begin{subfigure}{0.32\linewidth}
\caption{Manufacturing}
\includegraphics[width=1\linewidth]{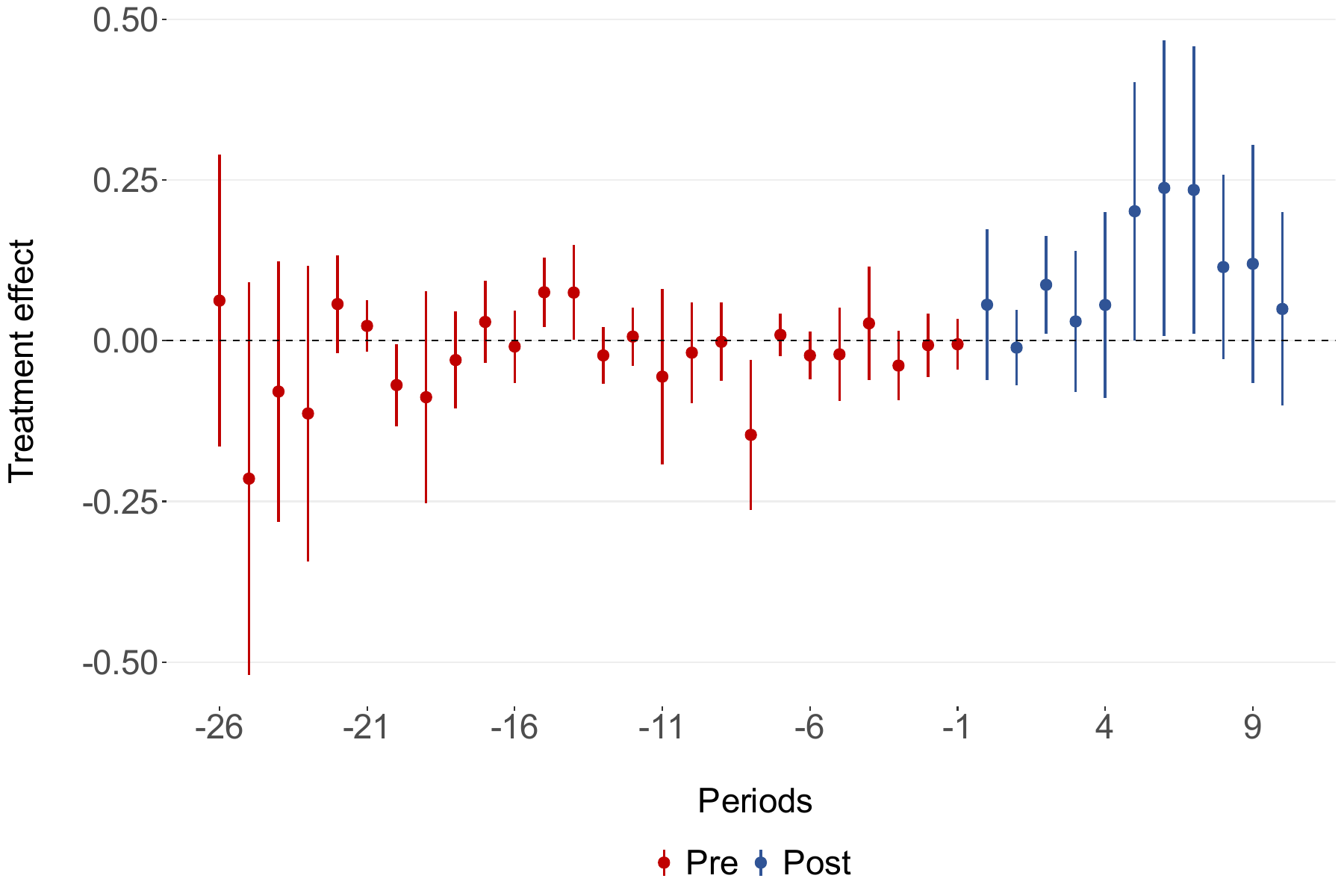}
\label{fig:vinculos_ativos_industria_dynamic_pandemia}
\end{subfigure}
\begin{subfigure}{0.32\linewidth}
\caption{Construction}
\includegraphics[width=1\linewidth]{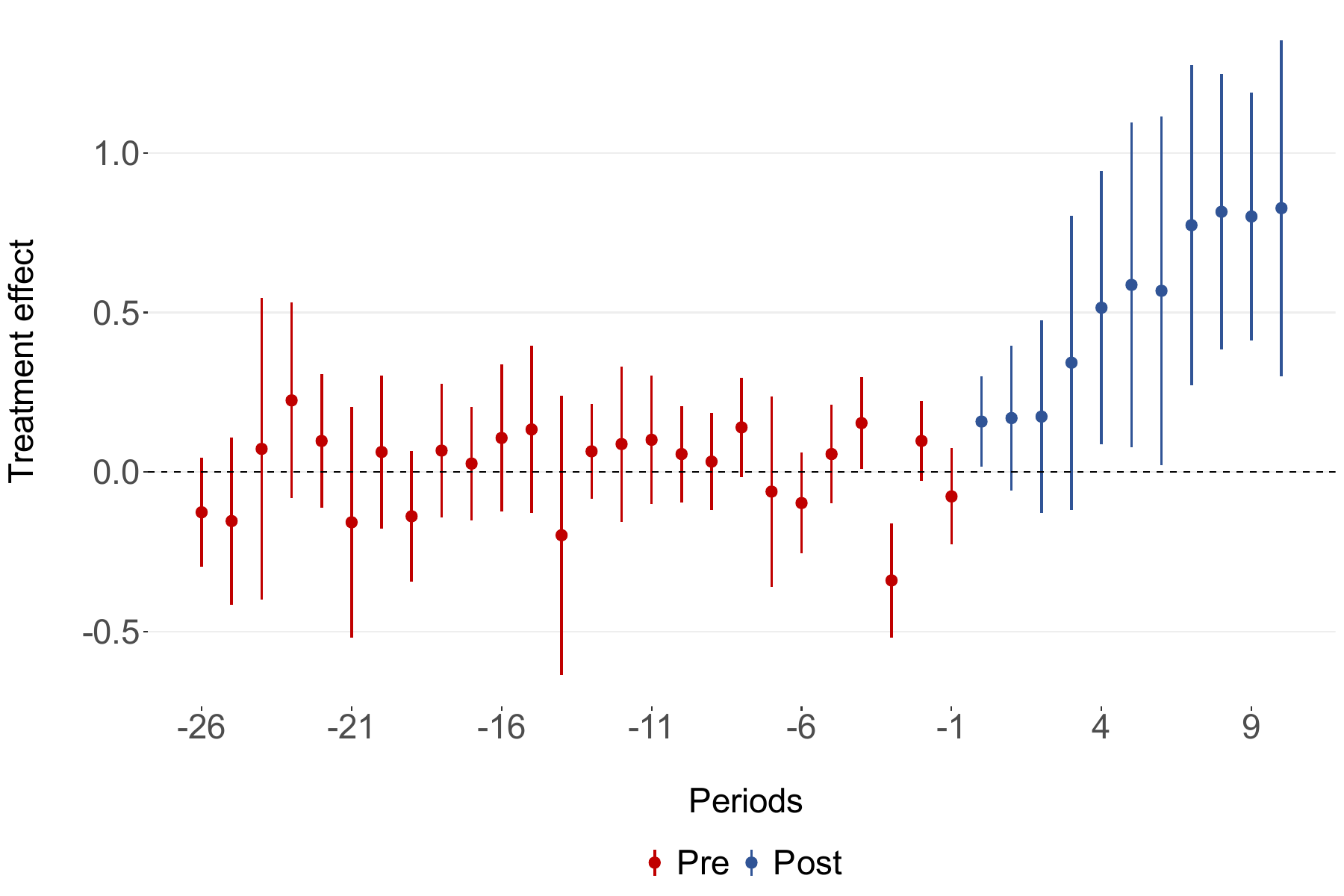}
\label{fig:vinculos_ativos_construcao_dynamic_pandemia}\end{subfigure}
\begin{subfigure}{0.32\linewidth}
\caption{Commerce}
\includegraphics[width=1\linewidth]{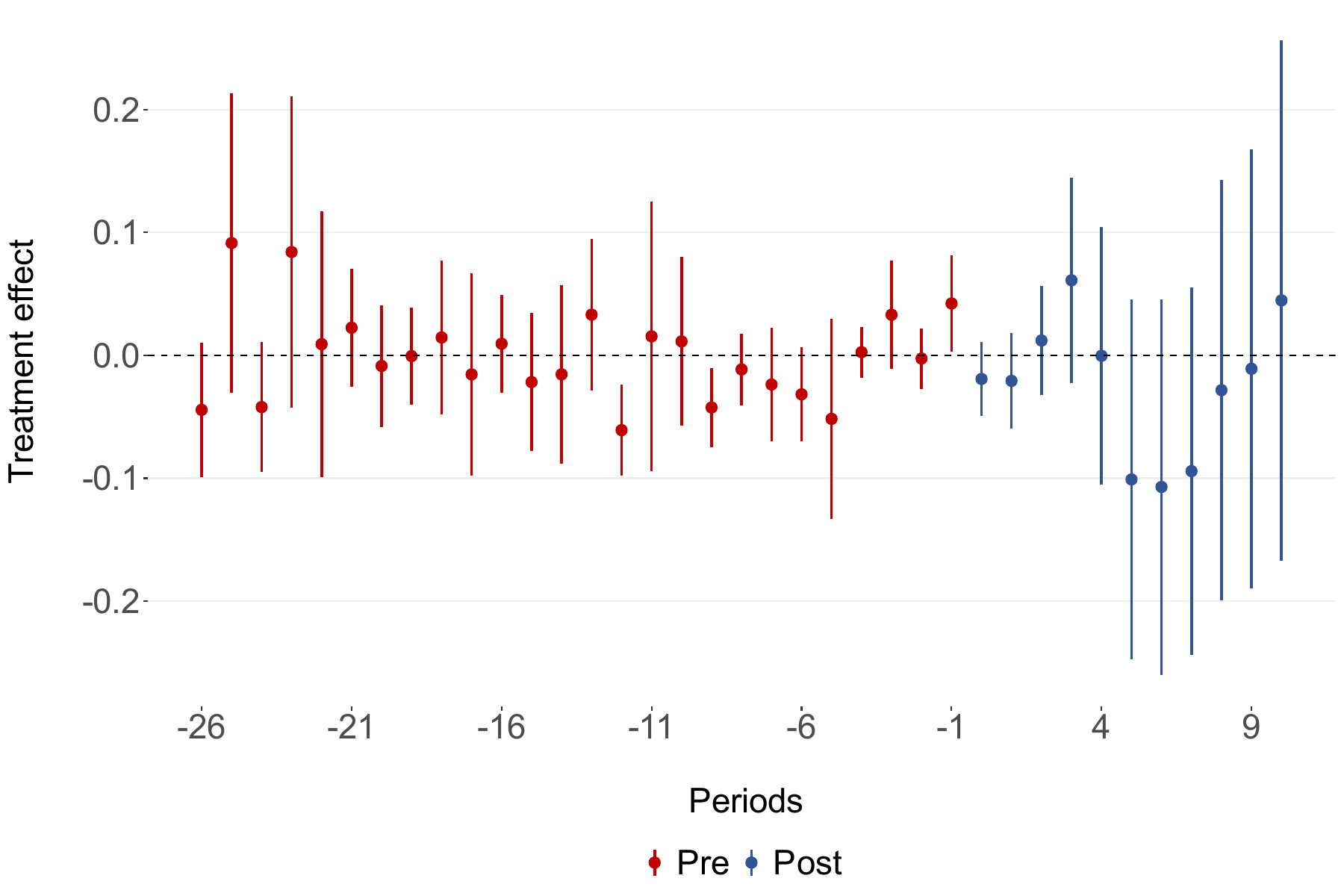}
\label{fig:vinculos_ativos_comercio_dynamic_pandemia}
\end{subfigure} \\
\begin{subfigure}{0.32\linewidth}
\caption{Services}
\includegraphics[width=1\linewidth]{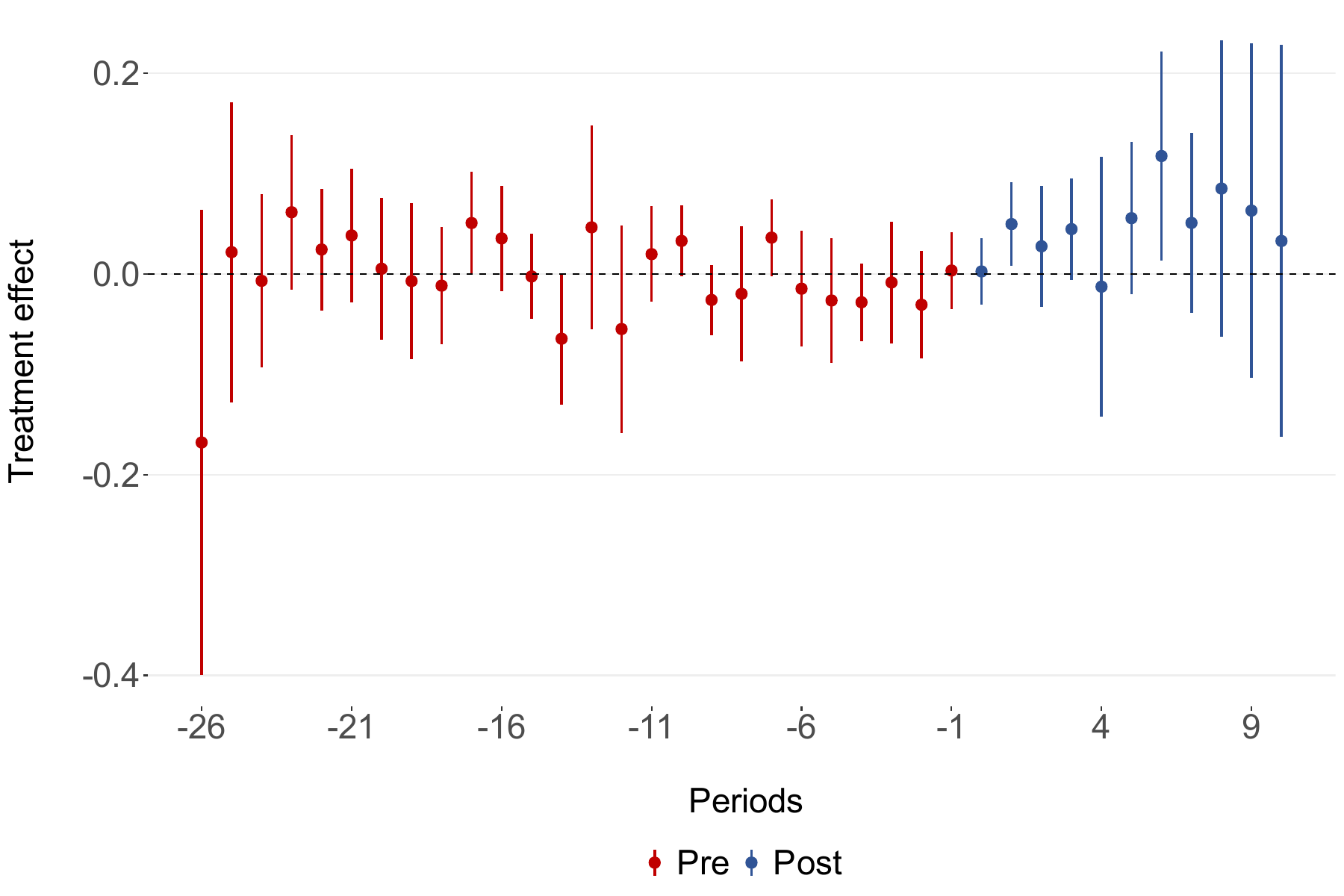}
\label{fig:vinculos_ativos_servicos_dynamic_pandemia}\end{subfigure}
\begin{subfigure}{0.32\linewidth}
\caption{Agriculture}
\includegraphics[width=1\linewidth]{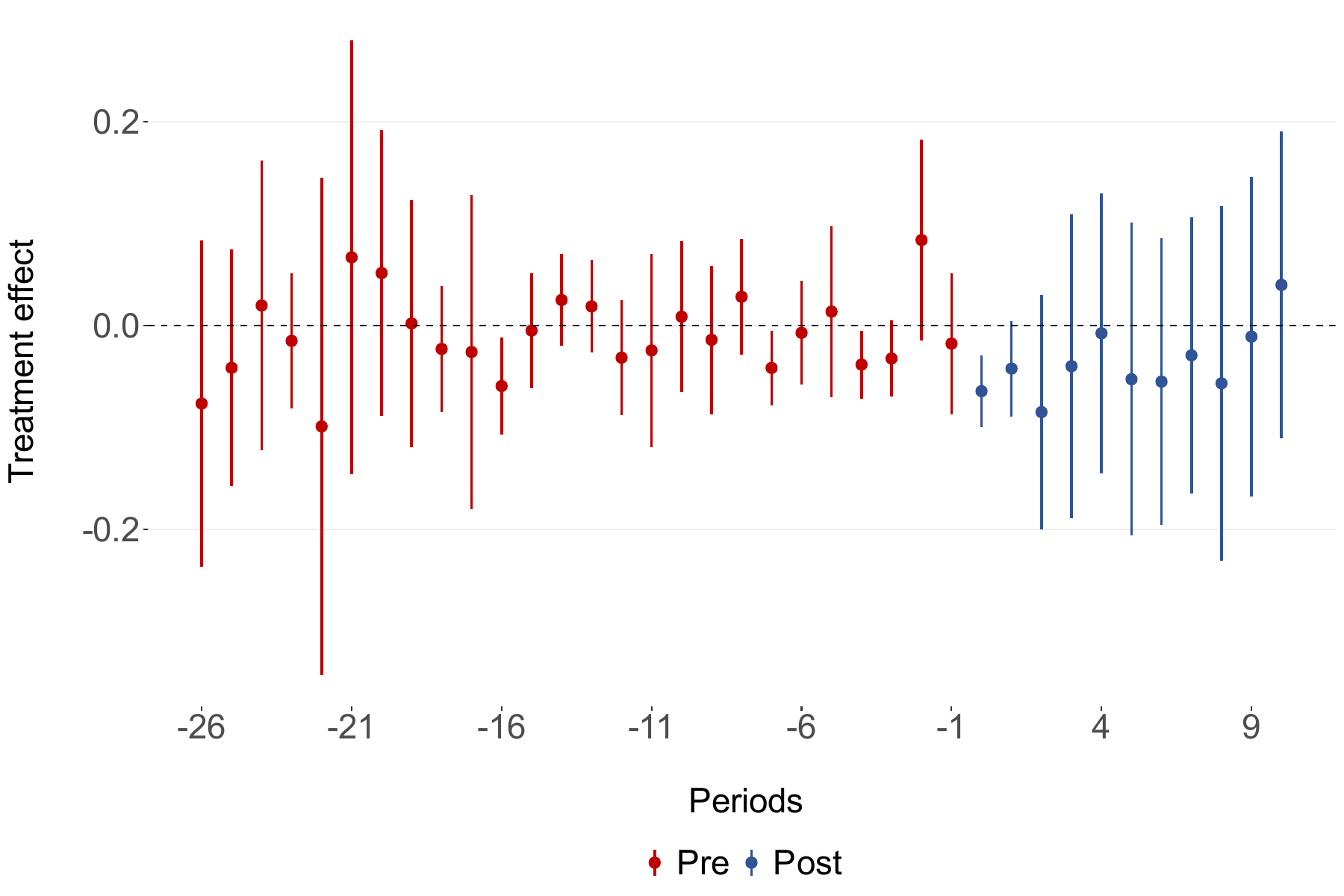}
\label{fig:vinculos_ativos_agricultura_dynamic_pandemia}
\end{subfigure}
\begin{subfigure}{0.32\linewidth}
\caption{Transportation}
\includegraphics[width=1\linewidth]{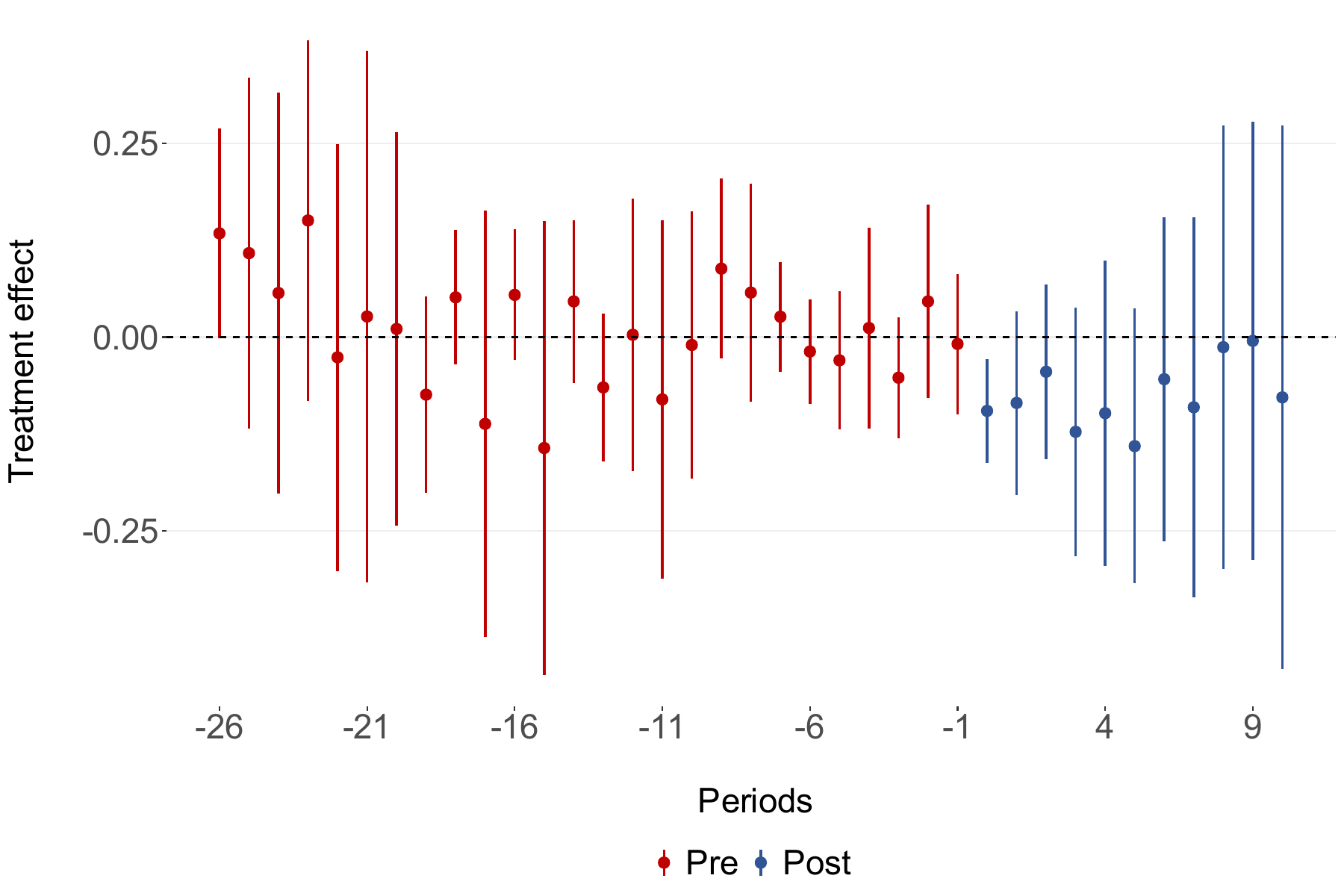}
\label{fig:vinculos_ativos_transporte_dynamic_pandemia}\end{subfigure} \\
\end{center}
\footnotesize{\textit{Notes:} Figure \ref{fig:vinculos_ativos_transporte_dynamic_pandemia} presents estimates of the average effect of adopting free public transit $e$ years after adoption across all municipalities that are ever observed to have taken the treatment for exactly $e$ periods (Equation \eqref{eq:treatment_effect_length_exposure}). The outcome variables are the natural logarithm of the formal employment in manufacturing in each municipality (Figure \ref{fig:vinculos_ativos_industria_dynamic_pandemia}), the natural logarithm of the formal employment in construction in each municipality (Figure \ref{fig:vinculos_ativos_construcao_dynamic_pandemia}), the natural logarithm of the formal employment in commerce in each municipality (Figure \ref{fig:vinculos_ativos_comercio_dynamic_pandemia}), the natural logarithm of the formal employment in services in each municipality (Figure \ref{fig:vinculos_ativos_servicos_dynamic_pandemia}), the natural logarithm of the formal employment in agriculture in each municipality (Figure \ref{fig:vinculos_ativos_agricultura_dynamic_pandemia}), and the natural logarithm of the formal employment in transportation in each municipality (Figure \ref{fig:vinculos_ativos_transporte_dynamic_pandemia}). These sectors are formally defined in Table \ref{tab:tab:sector_aggregation}. Vertical lines represent uniform 90\%-confidence intervals based on standard errors clusterized at the municipality level. These results are based on the doubly-robust estimator proposed by \cite{callaway2021difference} using a varying base period for the pre-treatment placebo estimates (in red). Post-treatment estimates are reported in blue.}
\end{figure}

\begin{figure}[htbp]
\begin{center}
\caption{Average Effects for each Length of Exposure to the Treatment \\ Outcome Variable: Average Formal Wage in each Municipality}
\label{fig:salario}
\includegraphics[width=1\linewidth]{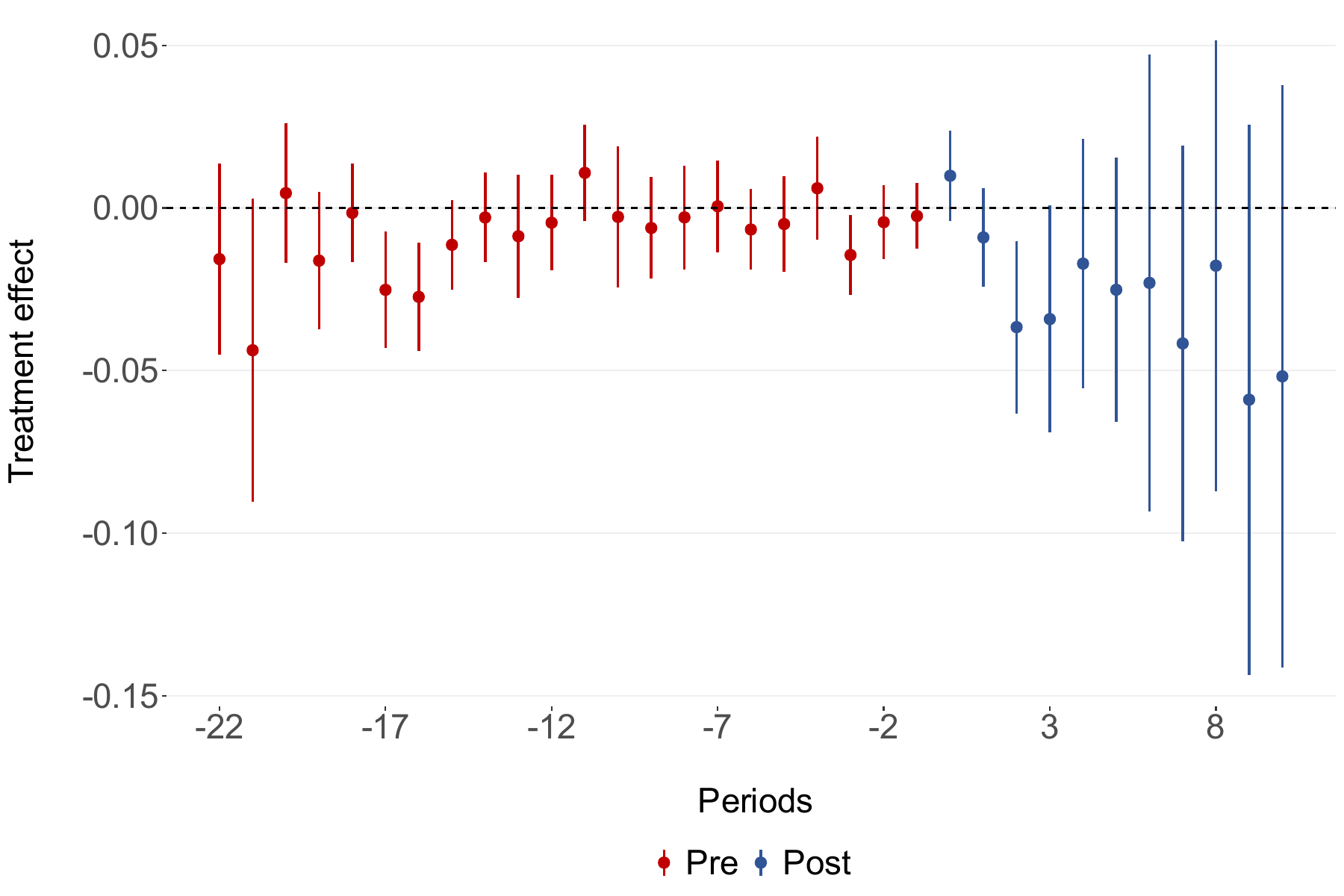}
\end{center}
\footnotesize{\textit{Notes:} Figure \ref{fig:salario} presents estimates of the average effect of adopting free public transit $e$ years after adoption across all municipalities that are ever observed to have taken the treatment for exactly $e$ periods (Equation \eqref{eq:treatment_effect_length_exposure}). The outcome variables are the natural logarithm of the average formal wage in each municipality. Vertical lines represent point-wise 90\%-confidence intervals based on standard errors clusterized at the municipality level. These results are based on the doubly-robust estimator proposed by \cite{callaway2021difference} using a varying base period for the pre-treatment placebo estimates (in red). Post-treatment estimates are reported in blue.}
\end{figure}

\begin{table}[H]
\begin{center}\caption{\label{tab:tab:sector_aggregation}Sector aggregation}
\begin{tabular}{lc}
\toprule \toprule
Aggregate sector & CNAE 1.0 section\\
\midrule
Manufacturing & C and D\\
Construction & F\\
Commerce & G\\
Services & H, J, K, L, M, N, O and P\\
Agriculture & A\\
Transportation & I\\
\bottomrule
\bottomrule
\end{tabular}
\end{center}
\footnotesize{\textit{Notes:} This table shows the sector aggregation based on CNAE 1.0 sections. The sections are arranged sequentially: Section A includes agriculture, livestock, forestry, and forest exploitation, as well as transportation, storage, and communications. Section C denotes extractive industries, while Section D represents manufacturing industries. For brevity, we combine both sectors into one and call it simply manufacturing. Section F pertains to construction, and Section G encompasses commerce and repair of motor vehicles, personal, and household items. Section H, J, K, L, M, N, O and P are combined into the services sector. Section H focuses on lodging and food, and Section J relates to financial intermediation. Further, Section K involves real estate activities, rentals, and services provided to companies, whereas Section L covers public administration, defense, and social security. Section M deals with education, followed by Section N, which encompasses health and social services. Section O comprises other collective, social, and personal services, and Section P denotes domestic services. Section I encompasses transportation services.}
\end{table}

\begin{table}[H]
\begin{center}
\caption{\label{tab:sample_selection} Number of observations in each step of the sample selection, by outcome}
\begin{tabular}{lccccc}
\toprule
\toprule
\multicolumn{1}{c}{} & \multicolumn{1}{c}{} & \multicolumn{4}{c}{Type of selection} \\
\cmidrule(l{3pt}r{3pt}){3-6}
 & Original & Control & Population & States & $>0$\\
\midrule
Emissions & 3800 & 3733 & 3730 & 2371 & 2361\\
Total employment & 3800 & 3733 & 3730 & 2371 & 2361\\
Stock of cars & 3800 & 3733 & 3730 & 2371 & 2369\\
\addlinespace[0.3em]
\multicolumn{6}{l}{\textit{Sectoral employment}}\\
\hspace{1em}Industry & 3800 & 3733 & 3730 & 2371 & 1982\\
\hspace{1em}Construction & 3800 & 3733 & 3730 & 2371 & 977\\
\hspace{1em}Trade & 3800 & 3733 & 3730 & 2371 & 2203\\
\hspace{1em}Services & 3800 & 3733 & 3730 & 2371 & 2325\\
\hspace{1em}Agriculture & 3800 & 3733 & 3730 & 2371 & 2162\\
\hspace{1em}Transportation & 3800 & 3733 & 3730 & 2371 & 2212\\
\addlinespace[0.3em]
\multicolumn{6}{l}{\textit{Sales by type of fuel}}\\
\hspace{1em}Gasoline & 3800 & 3733 & 3730 & 2371 & 2162\\
\hspace{1em}Ethanol & 3800 & 3733 & 3730 & 2371 & 1860\\
\bottomrule
\bottomrule
\end{tabular}
\end{center}
\footnotesize{\textit{Notes:} This table illustrates the sample selection procedure in terms of number of observations. All datasets begin with a number of 3,800 MCA. The column ``Control'' shows the number of observations that left after removing from the control group all MCA that implement any type of free public transport or subsidy. The column ``Population'' denotes the number of observations left after removing all the MCA that have population higher than the largest treated MCA. The column ``States'' shows the number of observations left after removing the Brazilian states that do not have at least one treated unit. Finally, column ``$>0$'' shows the final number of MCA in the sample used for the estimation of treatment effects, where we removed outcome-specific MCA that presented at least one observation equal to zero. Note: there were only 14 MCA with non-zero link values in 2022, a problem in the original dataset, so we present the value of MCA up until 2021, which is the one used in the regressions.}
\end{table}

\begin{table}[H]
\begin{center}
\caption{\label{tab:descriptive_statistics}Descriptive statistics of the MCA baseline characteristics}
\begin{tabular}{lcccc}
\toprule
\toprule
\multicolumn{1}{c}{} & \multicolumn{2}{c}{Treated} & \multicolumn{2}{c}{Control} \\
\cmidrule(l{3pt}r{3pt}){2-3} \cmidrule(l{3pt}r{3pt}){4-5}
 & Mean & Std. Deviation & Mean & Std. Deviation\\
\midrule
Emissions & 1646.91 & 8881.76 & 279.1 & 640.38\\
Total of jobs & 5382 & 11312 & 2729 & 8998\\
Stock of cars & 6546 & 9076 & 3994 & 11347\\
Mean Wage & 492.48 & 169.66 & 392.34 & 171.99\\
Fuel sales & 8640 & 20587 & 4415 & 11741\\
Total population & 58194 & 140602 & 28435 & 68434\\
\addlinespace[0.3em]
\multicolumn{5}{l}{\textit{Employment composition}}\\
\hspace{1em}Manufacturing & 0.27 & 0.20 & 0.23 & 0.19\\
\hspace{1em}Construction & 0.02 & 0.02 & 0.04 & 0.06\\
\hspace{1em}Commerce & 0.11 & 0.07 & 0.11 & 0.08\\
\hspace{1em}Services & 0.38 & 0.17 & 0.45 & 0.23\\
\hspace{1em}Agriculture & 0.13 & 0.12 & 0.18 & 0.17\\
\hspace{1em}Transportation & 0.04 & 0.05 & 0.02 & 0.04\\
\hspace{1em}Other sectors & 0.06 & 0.09 & 0.05 & 0.07\\
\addlinespace[0.3em]
\multicolumn{5}{l}{\textit{Fuel sales composition}}\\
\hspace{1em}Gasoline sales & 0.55 & 0.08 & 0.57 & 0.11\\
\hspace{1em}Ethanol sales & 0.45 & 0.08 & 0.44 & 0.1\\
\addlinespace[0.3em]
\multicolumn{5}{l}{\textit{Population composition}}\\
\hspace{1em}Urban population & 0.75 & 0.18 & 0.54 & 0.22\\
\hspace{1em}Rural population & 0.25 & 0.18 & 0.46 & 0.22\\
\bottomrule
\bottomrule
\end{tabular}
\end{center}
\footnotesize{\textit{Notes:} This table provides an overview of the baseline characteristics of the MCA by treatment group, with all statistics referencing the initial year available in the dataset. As a result, emissions data pertains to 1970, total job and wage figures to 1985, car stock to 2002, fuel sales (including composition) to 1990 and employment composition shares for each sector reflect the scenario in 1994. The exception are population statistics. Despite the availability of older data, we opted to reference the most recent Brazilian Census preceding the treatment of the first unit, i.e., we use the 1991 Brazilian Census. Emissions are expressed in $10^3$ tons of CO$_2$-equivalent.}
\end{table}

\begin{table}[htbp]
\begin{center}
\caption{\label{tab:overall_att_estimates_with_controls} Average Treatment Effects Across Different Model Specifications}
\begin{tabular}{lcc}
\toprule
\toprule
 & Employment & GHG Emissions\\
 & \footnotesize{(1)} & \footnotesize{(2)} \\
\midrule
No controls & $0.032^{***}$ & $-0.041^{*}$ \\
 & (0.011) & (0.021) \\
Population & $0.034^{***}$ & $-0.039^{*}$ \\
 & (0.011) & (0.02) \\
Urban Share & $0.039^{***}$ & $-0.067^{**}$ \\
 & (0.011) & (0.028) \\
Per Capita Income & $0.037^{***}$ & $-0.053^{**}$ \\
 & (0.01) & (0.022) \\
Years of Schooling & $0.031^{***}$ & $-0.043^{**}$ \\
 & (0.01) & (0.02)  \\
All controls & $0.032^{***}$ & $-0.041^{**}$  \\
 & (0.011) & (0.023)  \\
\midrule
Units & 2361 & 2361 \\
Treated units & 57 & 57 \\
Groups & 20 & 20 \\
\bottomrule
\bottomrule
\end{tabular}
\end{center}
\footnotesize{\textit{Notes:} This table presents the estimates of the average effect of adopting free public transit across all municipalities that ever took treatment across different model specifications (Equation \eqref{eq:overall_treatment_effect}). The outcome variables are the natural logarithms of formal employment and greenhouse gas (GHG) emissions. The first model runs the event-study design without any control variables. The following models control for: AMC population, the share of the AMC population in an urban area, per capita income in the AMC, average years of schooling at the AMC, and all of the above. These results are based on the doubly-robust estimator proposed by \cite{callaway2021difference}. Standard errors are reported in parenthesis and are clustered at the municipality level. At the bottom, we also report the total number of municipalities in our samples, the number of treated municipalities and the number of adoption-year groups. Significance levels are denoted as follows: $^{***}p<0.01$; $^{**}p<0.05$; $^{*}p<0.1$.}
\end{table}

\begin{table}[H]
\begin{center}
\caption{\label{tab:att_estimates_seeg_ghg_employment_until2021} Average Effect of Adopting Free Public Transit on Employment Across Treated Municipalities (until 2021)}
\begin{tabular}{lc}
\toprule
\toprule
 & Employment \\
\midrule
Treatment effect & $0.037^{***}$ \\
 & (0.011) \\
\midrule
Treated units & 41\\
Groups & 19\\
Units & 2361\\
\bottomrule
\bottomrule
\end{tabular}
\end{center}
\footnotesize{\textit{Notes:} This table presents the estimates of the overall treatment effects on emissions and employment, defined as in Equation \eqref{eq:overall_treatment_effect}, for a restricted sample until 2021. This restriction is motivated by a small methodological change in RAIS that occurred in 2022. The estimation was conducted using the doubly-robust estimator proposed by \cite{callaway2021difference}. Standard errors, clustered at the unit level, are presented in parentheses. Additionally, the table displays the total number of units used in the estimation, along with the number of treated units and treated groups. Significance levels are denoted as follows: $^{***}p<0.01$; $^{**}p<0.05$; $^{*}p<0.1$.}
\end{table}

\begin{table}[htbp]
\begin{center}
\caption{\label{tab:salarios}Average Effect of Adopting Free Public Transit Across Treated Municipalities}
\begin{tabular}{lc}
\toprule
\toprule
 & Average wage\\
\midrule
Treatment effect & $-0.014^{}$\\
 & (0.014)\\
\midrule
 Units & 2368\\
Treated units & 54\\
Groups & 17\\
\bottomrule
\bottomrule
\end{tabular}
\end{center}
\footnotesize{\textit{Notes:} This table presents the estimates of the average effect of adopting free public transit across all municipalities that ever took treatment (Equation \eqref{eq:overall_treatment_effect}). The outcome variable is the natural logarithm of the average formal wage in each municipality. These results are based on the doubly-robust estimator proposed by \cite{callaway2021difference}. Standard errors are reported in parenthesis and are clustered at the municipality level. At the bottom, we also report the total number of municipalities in our samples, the number of treated municipalities and the number of adoption-year groups. Significance levels are denoted as follows: $^{***}p<0.01$; $^{**}p<0.05$; $^{*}p<0.1$.}
\end{table}

\bigskip

%**********************************************************
%\addcontentsline{toc}{section}{References}
%\nocitesec{*}
%\bibliographystylesec{aer}
%\bibliographysec{otherreference}

\end{document}